\documentclass[aps,prd,twocolumn,superscriptaddress,showpacs,nofootinbib,amssymb,notitlepage]{revtex4-1}
\pdfoutput=1
\usepackage[usenames,dvipsnames]{color}
\usepackage[colorlinks=true, linkcolor=Black, citecolor=Black, urlcolor=Black, filecolor=Black]{hyperref}
\usepackage{longtable}
\usepackage{soul}
\usepackage{epsfig}
\usepackage{amssymb}
\usepackage{amsmath}
\usepackage{graphicx}
\usepackage{verbatim}
\usepackage{xspace}
\usepackage{hyperref}
\usepackage{float}

\newcommand{\nc}{\newcommand}
\nc{\be}{\begin{equation}}
\nc{\ee}{\end{equation}}
\nc{\bea}{\begin{eqnarray}}
\nc{\eea}{\end{eqnarray}}

\begin{document}

\renewcommand{\arraystretch}{1.5}

\title{Dynamically vanishing Dirac neutrino mass from quantum scale symmetry}

\author{Astrid Eichhorn}
\email{eichhorn@cp3.sdu.dk}
\affiliation{CP3-Origins, University of Southern Denmark, Campusvej 55, DK-5230 Odense M, Denmark}

\author{Aaron Held}
\email{aaron.held@uni-jena.de}
\affiliation{
Theoretisch-Physikalisches Institut, Friedrich-Schiller-Universit\"at Jena,
Max-Wien-Platz 1, 07743 Jena, Germany
}
\affiliation{
The Princeton Gravity Initiative, Jadwin Hall, Princeton University,
Princeton, New Jersey 08544, U.S.
}

\begin{abstract}
We present a mechanism which drives Dirac neutrino masses to tiny values along the Renormalization Group flow, starting from an asymptotically safe ultraviolet completion of the third generation of the Standard Model including quantum gravity. At the same time, the mechanism produces a mass-splitting between the neutrino and the quark sector and also generates the mass splitting between top and bottom quark. The mechanism hinges on the hypercharges of the fermions and produces a tiny neutrino Yukawa coupling, because the right-handed neutrino is sterile and does not carry hypercharge.
\end{abstract}

\maketitle

\section{Why are neutrinos (almost) massless?}

Among the fundamental fermions in nature, neutrinos are special: they do not carry electric charge; we have only observed their left-handed chiral component; and, when the Standard Model of particle physics (SM) was constructed, they were assumed to be massless.

Even to date, neutrino masses have not been directly measured~\cite{ParticleDataGroup:2020ssz}.
However, the observation of flavor oscillations in solar \cite{Davis:1968cp,Cleveland:1998nv,Fukuda:1996sz,Hampel:1998xg,Fukuda:2002pe,Ahmad:2002jz,Altmann:2005ix,Abdurashitov:2009tn}, atmospheric \cite{Kamiokande-II:1988sxn,Casper:1990ac,Fukuda:1998mi, Ashie:2004mr}, reactor \cite{Eguchi:2002dm,Araki:2004mb} and accelerator neutrinos \cite{Ahn:2006zza,Michael:2006rx} constitutes compelling indirect evidence for neutrinos masses. 
Majorana masses (of the effective electron-neutrino flavor) are constrained to $\lesssim 1\;\text{eV}$ by direct experimental searches for neutrinoless double-$\beta$ decay~\cite{KamLAND-Zen:2016pfg, KATRIN:2019yun}
Dirac masses are not constrained by direct experimental data. Neutrino-oscillation experiments constrain the differences between neutrino masses to the $\sim$meV range, cf.~\cite{Gonzalez-Garcia:2012hef, Esteban:2020cvm} for best fits. Within the $\Lambda$CDM cosmological model, observations of the cosmic microwave background~\cite{Palanque-Delabrouille:2015pga, Aghanim:2016yuo, Planck:2018vyg} constrain the overall neutrino mass (Dirac or Majorana) to $\sum_j m_{\nu\,j} < 120\,\text{meV}$ (where $j$ labels the neutrino species).
\\

The SM gauge symmetry forbids Majorana mass terms for the left-handed neutrinos and Dirac masses require additional right-handed neutrinos. Hence, neutrino masses require an extension of the SM
by new degrees of freedom.
Right-handed sterile neutrinos constitute a minimal extension of the SM -- the $\nu$SM -- and they can interact with the left-handed lepton and the Higgs field through a Yukawa term. After electroweak symmetry breaking at $M_{\rm ew}$, Dirac neutrino masses
arise just like for all other SM fermions.

In the $\nu$SM, neutrino masses in the meV range translate into neutrino Yukawa couplings of the order of $y_\nu\sim10^{-13}$. In the literature, the value of $y_\nu$ is usually seen as unnaturally small, while the electron Yukawa coupling of the order $y_{e}\sim10^{-5}$ is usually deemed acceptable.
Thus, many beyond-SM scenarios contain a mechanism that generates small neutrino masses by introducing extra degrees of freedom beyond the $\nu$SM at some sufficiently high-energy scale $M_\text{NP}$, e.g., in type-I seesaw~\cite{Minkowski:1977sc,Yanagida:1979as,Glashow:1979nm,Gell-Mann:1979vob,Mohapatra:1979ia}, type-II seesaw \cite{Magg:1980ut,Lazarides:1980nt,Mohapatra:1980yp} or type-III seesaw \cite{Foot:1988aq}, see, e.g., \cite{King:2003jb,Mohapatra:2006gs,Xing:2020ald} for reviews. The smallness of neutrino masses is then tied to the ratio $M_\text{ew}/M_\text{NP}$.
\\

Here, we follow a somewhat different line of reasoning. We consider all free parameters of the $\nu$SM on an equal footing. Whether or not these free parameters are small or not is not a decisive factor in this line of reasoning.\footnote{Observations indeed point to nature being ``unnatural" in that several very small parameters exist which are not tied to new physics (at least at the current state of observations): the first is the accelerated late-time expansion of the universe, which is explained by a cosmological constant which is many orders of magnitude smaller than the Planck mass. The second is the Higgs mass, which is also several orders of magnitude smaller than the Planck mass. The last is the neutron electric dipole moment, which, in appropriate units, is much smaller than one.}
Instead, we investigate whether a fundamental model of particle physics and gravity can explain some of the free parameters of the $\nu$SM -- irrespective of whether or not they are small.
We ask whether a quantum scale-invariant regime due to an asymptotically safe fixed point can provide such a fundamental explanation of some of the free parameters, including the neutrino Yukawa coupling. We consider asymptotic safety of gravity and matter~\cite{Eichhorn:2018yfc,Eichhorn:2022jqj} as a promising candidate to achieve such an explanation, because it has been found to reduce the number of free parameters in the SM~\cite{Shaposhnikov:2009pv,
Harst:2011zx,
Eichhorn:2017ylw,
Eichhorn:2017lry,Eichhorn:2018whv}
and beyond~\cite{Eichhorn:2017muy,
Eichhorn:2017als,
Eichhorn:2019dhg,
Reichert:2019car,
Eichhorn:2020kca,
Eichhorn:2020sbo,
Hamada:2020vnf,
Kowalska:2020gie,
Kowalska:2020zve,
Held:2022hnw},
to dynamically generate small scales~\cite{Wetterich:2016uxm} and to generate mass-splittings, e.g., between top- and bottom-quark mass~\cite{Eichhorn:2018whv}.

\section{Mass from Charge}
The asymptotic-safety paradigm~\cite{Weinberg:1980gg} relies on interacting fixed points of the Renormalization Group (RG) flow. Such fixed points serve two purposes: first, they ultraviolet (UV) complete a theory; second, they reduce the number of free parameters of a theory to finitely many.
The theory is UV complete because the running of couplings with RG scale $k$ vanishes at an interacting fixed point, i.e., all beta functions are zero. Thus, Landau poles, which would prevent a UV completion, are absent from the RG flow.
The theory is predictive because the fixed-point regime can only be left along relevant (infrared (IR) repulsive) directions in the space of couplings. The IR values of these relevant couplings are the free parameters of the theory. All other couplings in the low-energy effective field theory are irrelevant (IR attractive) directions, i.e., calculable in terms of the relevant couplings.\\
Asymptotic safety shares these properties (UV completeness and predictivity) with asymptotic freedom: both correspond to UV completions in scale-invariant regimes, i.e., with vanishing beta functions. Further, in both, the IR physics only depends on a finite number of free parameters, the relevant couplings. The difference between asymptotic safety and asymptotic freedom lies in the set of relevant couplings: in asymptotic freedom, couplings with positive mass dimension are relevant; in addition, couplings with vanishing mass dimension can be marginally relevant if the lowest-order term in the beta function is negative. In asymptotic safety, quantum corrections to the scaling dimensions are present. Therefore, the set of relevant couplings is not determined by the mass dimensionality alone and needs to be determined by an explicit calculation of the beta functions.

The SM is neither asymptotically free nor asymptotically safe; therefore, a UV completion requires additional degrees of freedom that can generate the required zeros in the beta function. Intriguingly, gravity -- the only known fundamental force which is not included in the SM -- appears to provide the missing degrees of freedom. Starting from \cite{Reuter:1996cp}, compelling evidence for asymptotic safety has accumulated, see \cite{Eichhorn:2018yfc,Reichert:2020mja,Pawlowski:2020qer,Eichhorn:2022jqj} for recent reviews, \cite{Percacci:2017fkn,Reuter:2019byg} for textbooks and \cite{Bonanno:2020bil} for a critical discussion of the open challenges of the program, with the more recent works \cite{Draper:2020bop,Platania:2020knd,Fehre:2021eob} addressing a key open question -- unitarity. Most importantly for our purposes, an asymptotically safe fixed point exists in studies of quantum gravity minimally coupled to SM fields \cite{Dona:2013qba,Meibohm:2015twa,Biemans:2017zca,Alkofer:2018fxj,Wetterich:2019zdo,Sen:2021ffc}. 
The impact of interactions has been studied in \cite{
Narain:2009fy, 
Folkerts:2011jz,
Eichhorn:2011pc, 
Eichhorn:2012va, 
Percacci:2015wwa,
Eichhorn:2016vvy, 
Eichhorn:2017sok, 
Christiansen:2017gtg,
Christiansen:2017cxa, 
Eichhorn:2017sok, 
Eichhorn:2017eht,
Eichhorn:2017als, 
Eichhorn:2018nda,
Pawlowski:2018ixd,
Eichhorn:2018nda,
Eichhorn:2018ydy,
Eichhorn:2018akn,
Gies:2018jnv,
Eichhorn:2019yzm,
DeBrito:2019rrh,
Wetterich:2019rsn,
Wetterich:2019zdo,
Hamada:2020mug,
deBrito:2020dta,
Eichhorn:2020sbo,
Daas:2021abx, 
Gies:2021upb,
deBrito:2021pyi,
Laporte:2021kyp,
Ohta:2021bkc,
Eichhorn:2021qet} which provide good indications for the persistence of this fixed point. For our purposes, the most relevant piece of information is the following: the leading-order gravitational contribution to the beta functions of all gauge couplings and all Yukawa couplings in the SM can be written in terms of one parameter each, i.e., just two parameters in total. These enter the beta functions at the linear level in the respective SM couplings.\footnote{There is a potential confusion regarding the beta functions with gravitational contributions, namely, that the gravitational contributions do not impact the logarithmic running of couplings, see \cite{Donoghue:2019clr,Bonanno:2020bil}. For our purposes, this is unimportant -- what matters is whether or not couplings are made irrelevant by the gravitational contributions, such that their IR value can be predicted. This does not depend on whether or not the gravitational contribution leads to running in the sense of a logarithmic scale dependence.} These two parameters, $f_g$ and $f_y$, can be understood as parametrizing the gravitational fluctuation in asymptotic safety. 
Alternatively, they can be understood as parametrizing some unknown new physics which i) kicks in beyond the Planck scale, $k> M_\text{Planck}\approx 10^{19}\text{GeV}$, ii) is blind to internal symmetries, such that all Yukawa beta functions are modified by $f_y$ and all gauge beta functions are modified by $f_g$, iii) is antiscreening, i.e., $\beta_y = -f_y\, y + \mathcal{O}(y)^3$ and $\beta_g = -f_g\, g + \mathcal{O}(g^3)$ for all Yukawa couplings $y$ and gauge couplings $g$ with
\begin{align}
f_y&=
\begin{cases}
0,\quad\quad\,  k<M_{\rm Planck}\\
{\rm const}, \, \, k \geq M_{\rm Planck}
\end{cases},
\\
f_g&=
\begin{cases}
0,\quad\quad \, k<M_{\rm Planck}\\
 {\rm const}, \, \, k \geq M_{\rm Planck}
\end{cases}
\;.
\end{align}
For definiteness, we focus on a Planckian new-physics scale; however, a more general case substitutes $M_{\rm Planck}$ with a generic new-physics scale $M_{\rm NP}$.
\\

The leading gravitational contribution to the running of gauge and Yukawa couplings, $f_g$ and $f_y$ have been calculated with functional RG (FRG) techniques, see \cite{Daum:2009dn,Daum:2010bc,Harst:2011zx,Folkerts:2011jz,Christiansen:2017cxa,Eichhorn:2017lry,DeBrito:2019gdd} and \cite{Oda:2015sma,Eichhorn:2016esv,Eichhorn:2017eht,Hamada:2017rvn,Eichhorn:2017ylw,DeBrito:2019gdd,Eichhorn:2020sbo} respectively. 
The FRG \cite{Wetterich:1992yh,Morris:1993qb} provides a non-perturbative tool to extract the scale dependence of couplings and thereby obtain the effective dynamics at RG scale $k$. It is broadly applicable, from strongly-coupled fermion systems in condensed matter through universality classes in statistical physics to the non-perturbative regime of QCD and the understanding of quantum gravity, see ~\cite{Dupuis:2020fhh} for a recent review. It is particularly well-suited for systems characterized by dimensionful couplings, such as quantum gravity.

Both contributions ($f_g$ and $f_y$) depend on the gravitational fixed-point values, which are only determined with a comparatively large systematic uncertainty. Therefore, the exact values of $f_g$ and $f_y$ are both affected by systematic uncertainties. There are indications that under the impact of SM matter fields, the gravitational fixed-point values move into a regime, where $f_g>0$ and $f_y>0$ \cite{Eichhorn:2017ylw,Eichhorn:2018whv,Eichhorn:2020sbo} and also $f_{g,y}\ll~1$.

It is worth pointing out a subtlety that has led to some confusion in the literature: when $f_g$ and $f_y$ are calculated while treating the gravitational couplings as free parameters, they may vanish in some schemes \cite{Robinson:2005fj, Toms:2007sk,Ebert:2007gf,Anber:2010uj,Ellis:2010rw,Toms:2011zza,Felipe:2011rs,Narain:2012te}. In~\cite{deBrito:2022vbr} it has been pointed out that $f_g$ and $f_y$ are universal quantities when evaluated at the gravitational fixed point and that they do not vanish in that case.

In the following, we do not rely on explicit calculations of $f_g$ and $f_y$. As detailed below, the predictive power of an asymptotically safe fixed point can be strong enough to fix $f_g$ and $f_y$ in terms of two observations and still make further non-trivial predictions.

In contrast to the scheme dependence of $f_g$ and $f_y$, all the leading-order contributions arising from self-interactions within the SM only involve marginal couplings and are thus RG scheme-independent. The functional RG method, which is an appropriate method to use for the gravitational contribution, therefore leads to the well-known one-loop result for the matter contribution.\\

In \cite{Eichhorn:2018whv}, we proposed a mechanism by which the above two-parameter UV-completion of the SM can generate a mass splitting between up- and down-type quarks due to their distinct hypercharges. 

For large enough $f_g$, the antiscreening new-physics contribution to the running of the gauge couplings\footnote{If we do not specify the SM lepton charges, cf.~Sec.~\ref{sec:varyCharges}, $b_{0, \, Y}=\frac{1}{3}Y_\text{tot}$ with $Y_\text{tot} = 13 + 6 (Y_L^2+Y_{\nu}^2+Y_{\tau}^2)$ where $Y_L$, $Y_\tau$, and $Y_\nu$ respectively denote the hypercharges of the left-handed doublet, the right-handed $\tau$ lepton, and the right-handed neutrino. In the SM, $Y_L = -1/2$, $Y_\tau = 1$, and $Y_\nu=0$ such that $b_{0, \, Y}=41/6$.}, 
\begin{align}
\label{eq:betasg}
	\beta_{g_i} &= k \partial_k\, g_i(k)= \frac{b_{0,\,i}\,g_i^3}{16\pi^2}- f_g\, g_i,
	\\\notag
	\text{with}\quad
	b_{0,\, 3}&=-7,\, 
	b_{0,\,2}=-\frac{19}{6},\, 
	b_{0, \, Y}= \frac{41}{6},
\end{align}
overcomes the screening fermionic fluctuations and induces a UV-completion of the U(1) hypercharge sector \cite{Robinson:2005fj,Harst:2011zx, Eichhorn:2017lry}. When evaluated at the gravitational fixed point, as first done in \cite{Harst:2011zx}, $f_g$ becomes a critical exponent which is universal, see \cite{deBrito:2022vbr}.
While the non-Abelian gauge couplings remain asymptotically free, the hypercharge coupling is either asymptotically free or safe. In the latter case, its value follows as a prediction from the microscopic UV-scaling regime. Requiring that the fine-structure constant $\alpha_Y = g_Y^2/(4\pi)$ takes its observed IR value fixes $f_g\approx
9.7\times10^{-3}
$~\cite{Eichhorn:2017lry,Eichhorn:2018whv}.
For any value\footnote{This value may seem small, but note that the non-gravitational contributions in the $\beta$-functions come with a relative factor of $1/16\pi^2$. Pulling out a corresponding $1/16\pi^2$ would result in $f_g \sim \mathcal{O}(1)$.} $f_g>
9.7\times10^{-3}
$, the hypercharge gauge coupling is asymptotically free given its IR value as input. Here, we will focus on the case of asymptotically safe hypercharge coupling, $\frac{g_{Y,\,\ast}^2}{16\pi^2} = \frac{6}{41}f_g$ and $g_{2,\,\ast}^2 = g_{3,\,\ast}^2 = 0$, because this scenario i) lies well within the systematic uncertainties affecting the gravitational fixed-point values~\cite{Eichhorn:2017lry}, ii) is more predictive and iii) can lead to interesting structures in the Yukawa sector: in the presence of a flavor-blind contribution $f_y$ and for asymptotically free non-Abelian gauge couplings, the hypercharge fluctuations are the only ones to distinguish different flavors. Therefore, an explanation of why the top quark is so much heavier than the bottom quark and why the neutrino Yukawa coupling is generically tiny must hinge on $g_Y$ being nonzero.
\\
\begin{table*}[t]
\centering
\begin{tabular}{c|c|c|c||c|c|c|c||c|c|c|c}
	\multicolumn{4}{c||}{IR physics} & 
	\multicolumn{4}{c||}{fixed point} &
	\multicolumn{4}{c}{critical exponents}
\\
	$y_{t,\,\text{IR}}$ &
	$y_{b,\,\text{IR}}$ &
	$y_{\tau,\,\text{IR}}$ &
	$y_{\nu,\,\text{IR}}$ &
	$\frac{y_{t,\,\ast}^2}{16\pi^2}$ & 
	$\frac{y_{b,\,\ast}^2}{16\pi^2}$ & 
	$\frac{y_{\tau,\,\ast}^2}{16\pi^2}$ & 
	$\frac{y_{\nu,\,\ast}^2}{16\pi^2}$ & 
	$\theta_{{\nu}}$ & 
	$\theta_{t}$ & 
	$\theta_{b}$ & 
	$\theta_{\tau}$
\\\hline\hline
 	$\gtrsim 0.97$ (P) & 
 	$\checkmark$  (P) &
 	$\checkmark$ &
 	$\rightarrow0$  (P) &
 	$\frac{f_y}{6}+\frac{23 f_g}{492}$ & 
 	$ \frac{f_y}{6}-\frac{f_g}{492} $ & 
 	$ 0 $ & 
 	$ 0 $ & 
 	$-\frac{f_g}{41}$ & 
 	\multicolumn{2}{c|}{$\frac{\left(-33 f_g-246 f_y \pm A\right)}{164}$} &
 	$\frac{17f_g}{41}$ 
\\\hline
	$\checkmark$ &
 	$\checkmark$ &
 	$\checkmark$ &
 	$\checkmark$ &
	$ 0 $ & 
	$ 0 $ & 
	$ 0 $ & 
	$ 0 $ &
	$f_y+\frac{9f_g}{82}$ & 
	$f_y+\frac{17f_g}{82}$ & 
	$f_y+\frac{5f_g}{82}$ & 
	$f_y+\frac{45f_g}{82}$ 
\end{tabular}
\caption{\label{tab:fps}
We specify to SM charges and list the viable asymptotically free (bottom line) as well as the phenomenologically relevant interacting fixed point of the set of Eqs.~\eqref{eq:betayt3}-\eqref{eq:betaynu3}, along with the respective imprint on IR physics. All other fixed points are phenomenologically excluded or less predictive
(see main text). 
We also list the corresponding critical exponents, where $A=\sqrt{1804 f_g f_y+1273 f_g^2+6724 f_y^2}$. 
Whenever critical exponents extend over two columns, the respective eigendirection has overlap with both couplings. Otherwise, the respective eigendirection is aligned with the coupling. A `$\checkmark$' indicates that the measured IR value can be obtained. A `(P)' indicates that the IR value is a prediction that follows from the UV fixed point.
}
\end{table*}

In the Yukawa sector, we focus on a single generation of quarks and leptons, which we extend by a right-handed neutrino which is a gauge singlet, such that a Yukawa coupling between the left-handed SU(2) lepton doublet, the Higgs field and the right-handed neutrino is allowed.
The corresponding beta-functions read
\begin{align}
	\beta_{y_t}=&\frac{y_t}{16\pi^2}\left[ \frac{9}{2}y_t^2 + \frac{3}{2}y_b^2 + y^2_{\tau}+ y^2_{\nu} -\frac{17}{12} g_Y^2\right]-f_y\, y_t,
	\label{eq:betayt3}
	\\[0.5em]
	\beta_{y_b}=&\frac{y_b}{16\pi^2}\left[ \frac{9}{2}y_b^2 + \frac{3}{2}y_t^2 + y^2_{\tau}+ y^2_{\nu} -\frac{5}{12} g_Y^2\right]-f_y\, y_b,
	\label{eq:betayb3}
	\\[0.5em]
	\beta_{y_{\tau}}=&\frac{y_{\tau}}{16\pi^2}\left[3y_b^2 + 3y_t^2 + \frac{5}{2}y^2_{\tau}-\frac{1}{2} y^2_{\nu} - 3\,g_Y^2(Y_L^2 + Y_{\tau}^2)\right]
	\notag\\&
	-f_y\, y_{\tau},
	\label{eq:betaytau3}
	\\[0.5em]
	\beta_{y_{\nu}}=&\frac{y_{\nu}}{16\pi^2}\left[3y_b^2 + 3y_t^2 + \frac{5}{2}y^2_{\nu}-\frac{1}{2} y^2_{\tau} - 3\,g_Y^2(Y_L^2 + Y_{\nu}^2)\right]
	\notag\\&
	-f_y\, y_{\nu}.
	\label{eq:betaynu3}
\end{align}
In the quark sector, we have fixed the hypercharges to their SM values.
In the lepton sector, we denote the hypercharges of the left-handed lepton-doublet, the right-handed $\tau$-lepton and the right-handed neutrino by $Y_L$, $Y_\tau$, and $Y_{\nu}$, respectively. We do so to demonstrate that the mechanism for dynamically vanishing Dirac neutrino masses is connected to the specific charge assignment realized in the SM. Later, we will also set the lepton hypercharges to their SM values, i.e., $Y_L = -1/2$, $Y_\tau = 1$, and $Y_\nu=0$. 

An asymptotically free UV-completion exists whenever in all four beta-functions the antiscreening contributions (negative signs) from gauge- and new-physics fluctuations overcome the screening ones (positive signs) from matter self-interactions. Fixed points with non-vanishing Yukawa couplings also exist, when there is an antiscreening contribution to balance out the screening one.

For a set of equally charged fermions, the respective fixed-point structure is symmetric under exchanges of any of the fermions. Both the coupling to a Higgs doublet as well as the non-trivial charge assignment in the SM break this exchange symmetry. At fixed $f_y$, the available set of RG fixed points can thus imprint a splitting of the different Yukawa couplings at the UV fixed point, which persists along the RG flow and generates a mass splitting in the IR. 

This mass-splitting mechanism is at work in \cite{Eichhorn:2018whv}. We review it here, before asking whether the corresponding UV fixed point generates an interesting phenomenology for Dirac neutrino masses. If we set the hypercharge coupling to its fixed-point value, $g_{Y} = g_{Y\, \ast} = \sqrt{\frac{6 \cdot 16 \pi^2}{41}f_g}$, there are two antiscreening contributions in Eq.~\eqref{eq:betayt3} and Eq.~\eqref{eq:betayb3}. These differ in size, because the top quark carries a larger hypercharge than the bottom quark. In \cite{Eichhorn:2018whv}, it was thus found that there is a simple fixed-point relation between gauge coupling, top and bottom Yukawa coupling that reads
\be
y_{t\, \ast}^2 - y_{b\, \ast}^2 =\frac{1}{3}g_{Y\, \ast}^2\;. \label{eq:tbyFPrelation}
\ee
This relation implies that, as soon as the gauge coupling is set to its interacting fixed point, a mass-splitting between the top and bottom quark is generated, with the top quark being significantly heavier than the bottom quark. In \cite{Eichhorn:2018whv} it was found that the corresponding IR values come out in the vicinity of SM values: Choosing $f_g= 9.7\times10^{-3}
$ and $f_{y}= 1.188 \cdot 10^{-4}$ yields a hypercharge coupling and bottom Yukawa coupling in agreement with experimental values and a top quark that is somewhat too heavy (about 178 GeV)\footnote{This value was obtained in \cite{Eichhorn:2018whv} using 3-loop matching of running mass and pole mass, cf.~\cite[Eq.~(59.20)]{ParticleDataGroup:2020ssz}. Alternatively, a comparison to experimental data would require us to follow the functional RG flow throughout the electroweak phase transition, see~Sec.~\ref{sec:discussion}.}. The extension to three generations of quarks with a CKM mixing matrix was studied in \cite{Alkofer:2020vtb}. Because CKM matrix elements are essentially constant at a near-diagonal CKM matrix over a huge range of scales, the analysis in \cite{Eichhorn:2018whv} remains valid except for a deep-UV regime at far trans-Planckian scales where cross-over flows between different fixed-point regimes may start to play a role~\cite{Alkofer:2020vtb}.

Now we are ready to include the tau lepton and the tau-neutrino in our analysis and ask whether there is an extension of the fixed point in Eq.~\eqref{eq:tbyFPrelation} that provides an interesting phenomenology for the leptons.

\section{Free-Yukawa and safe-Yukawa fixed points}
In the Yukawa sector, most of the fixed points of Eq.~\eqref{eq:betayt3}-\eqref{eq:betaynu3} can be excluded as a fundamental UV completion because they are complex-valued or make either of the following predictions i) the top quark is heavier than 180~GeV, ii) the tau lepton is heavier than the top quark, or iii) the bottom quark has vanishing mass.
\\
Thereby, we identify the phenomenologically relevant fixed points, cf.~Tab.~\ref{tab:fps}.
The first one is the fixed point at which all Yukawa couplings vanish. 
We refer to it as the \textit{free-Yukawa fixed point}. It only makes a prediction for $g_Y$ and not for any of the Yukawa couplings because all four Yukawa couplings are relevant.
The second one is a partially interacting fixed point which extends the fixed point with $y_{t,\,\ast}\neq 0$ and $y_{b,\,\ast}\neq 0$ discussed in \cite{Eichhorn:2018whv}. We refer to it as the \textit{interacting-Yukawa fixed point}. It was first discussed in \cite{Held:2019vmi} and is given by 
\begin{align}
	\label{eq:interacting-FP}
	\frac{y_{t,\,\ast}^2}{16\pi^2} &= \frac{23 f_g}{24 Y_{\text{tot}}}+\frac{f_y}{6}\,,
	\notag\\
	\frac{y_{b,\,\ast}^2}{16\pi^2} &= \frac{f_y}{6}-\frac{f_g}{24 Y_{\text{tot}}}\,,
	\notag\\[0.5em]
	y_{\tau,\,\ast}^2 &= y_{\nu,\,\ast}^2 = 0\,,
\end{align}
with critical exponents
\begin{align}
	\label{eq:interacting-FP-CES}
	\theta_{t/b} &= \frac{-33 f_g-12 f_y Y_{\text{tot}}\pm A}{16 Y_{\text{tot}}}\,,
	\notag\\
	\theta_{\tau} &= \frac{f_g \left(36 Y_L^2+36 Y_{\tau }^2-11\right)}{4 Y_{\text{tot}}}\,,
	\notag\\
	\theta_{{\nu}} &= \frac{f_g \left(36 Y_L^2+36 Y_{\nu}^2-11\right)}{4 Y_{\text{tot}}}\,,
\end{align}
where $A = \sqrt{88 f_g f_y Y_{\text{tot}}+1273 f_g^2+16 f_y^2 Y_{\text{tot}}^2}$ and $Y_\text{tot} = 13 + 6 (Y_L^2+Y_{\nu}^2+Y_{\tau}^2)$.
The interacting-Yukawa fixed point exhibits IR-attractive (irrelevant) directions along which $g_Y$, $y_t$, and $y_b$ follow as predictions. This not only fixes the two new-physics parameters $f_g$ and $f_y$, but also ties the mass-difference of charged quarks to their charge difference~\cite{Eichhorn:2018whv}. The present extension to the lepton sector results in an additional IR-attractive direction which enforces $y_\nu=0$ and thus predicts a vanishing Dirac neutrino mass. In contrast, $y_\tau$ corresponds to an IR-repulsive (relevant) direction and a non-vanishing tau-lepton mass can arise in the IR.

There is one other interacting fixed point which can provide a phenomenologically viable fundamental UV-completion, namely the one at which $y_{t,\,\ast}\neq0$ is the only non-vanishing Yukawa coupling. It is less predictive in comparison to the interacting-Yukawa fixed point listed in Tab.~\ref{tab:fps} and has the same qualitative consequences for the neutrino Yukawa coupling.

For the phenomenology of the interacting-Yukawa fixed point, it is crucial that $\theta_{\nu}<0$, which we can trace back to the hypercharge assignment of the right-handed neutrino. Further, we discuss why a fully interacting fixed point, at which all four Yukawa couplings are non-zero, does actually not exist and thus there can never be more than three predictions coming out of the Yukawa sector of our model. We can trace this second property back to the coupling between the Higgs field and the leptons.

\subsection{Lepton charges and the Dirac neutrino mass}
\label{sec:varyCharges}
\begin{figure}[t]
\centering
\includegraphics[width=0.9\linewidth]{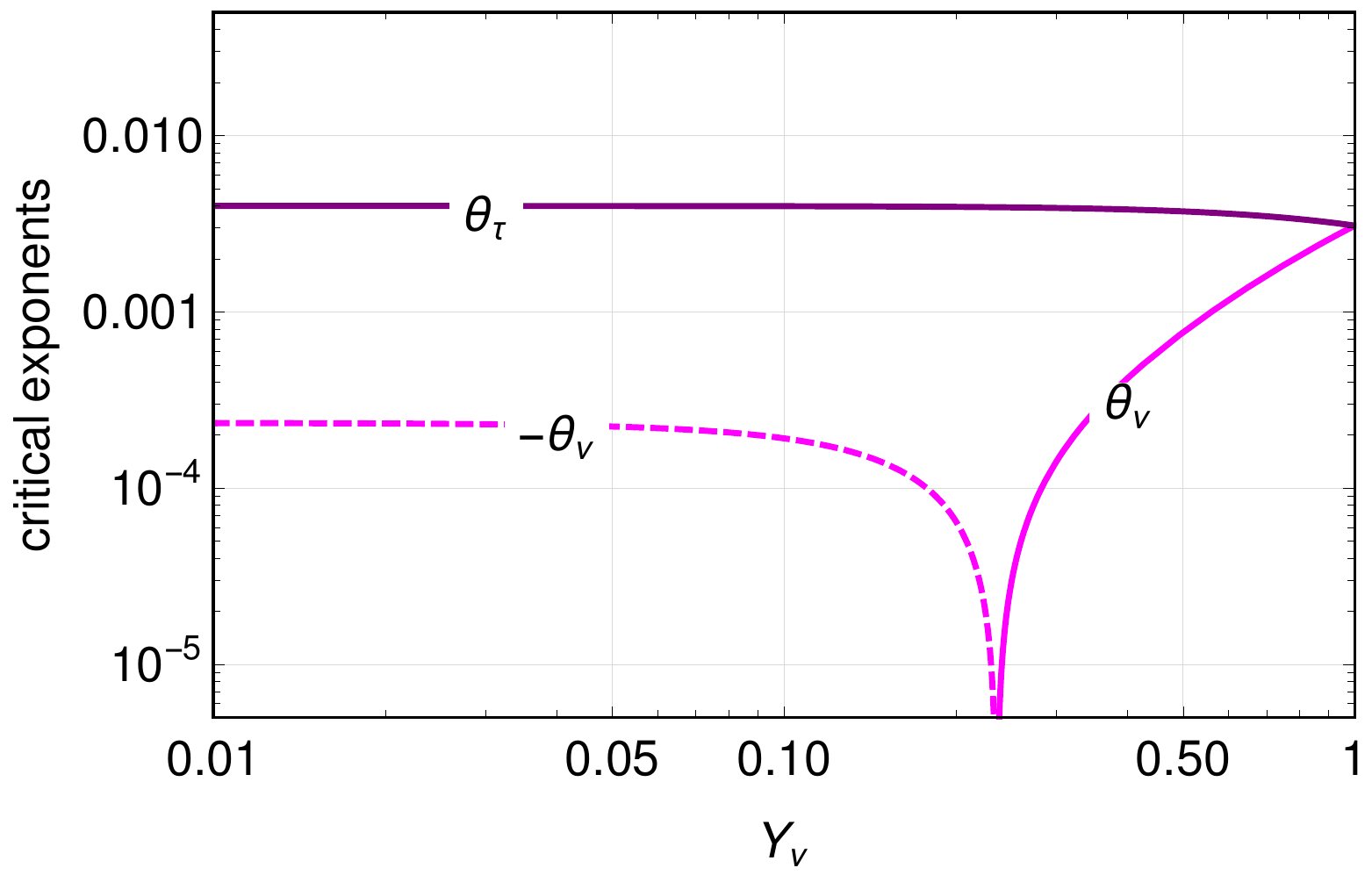}
\caption{\label{fig:thetasAndRunning}
We show $\theta_\tau$ (darker upper line) and $\theta_{\nu}$ (lower line) as a function of the neutrino hypercharge $Y_{\nu}$ with dashed (continuous) lines indicating negative (positive) $\theta_i$. For a sterile right-handed neutrino, i.e., $Y_{\nu}=0$, $\theta_{\nu}<0$.
}
\end{figure}

At the interacting-Yukawa fixed point, the critical exponents in the lepton sector are independent of $f_y$, cf.~Eq.~\eqref{eq:interacting-FP-CES}. Given that $f_g>0$, their signs are determined only by the corresponding hypercharges $Y_L$, $Y_\tau$, and $Y_{\nu}$. Away from SM charge assignments, either sign can be reached for both critical exponents. When we specify to SM charges, i.e., $Y_L=1/2$, $Y_\tau=-1$, and $Y_{\nu}=0$, the fixed point is IR-attractive in the neutrino and IR-repulsive in the tau-Yukawa coupling.
Thus, a non-vanishing $y_\tau$ can emerge from the UV-completion, even though $y_{\tau,\,\ast}=0$ at the fixed point. In particular, the IR-value which yields the measured tau-lepton mass can be reached.
In contrast, $y_{\nu}$ is dynamically driven to $y_{\nu,\,\ast}=0$, not just in the fixed-point regime, but at all scales. Consequently, the Dirac mass of the $\tau$-neutrino must vanish. In contrast, if $Y_{\nu}$ was not zero, but larger than $Y_{\nu\, \rm crit}^2 = \frac{1}{18}$, the neutrino Yukawa coupling would be relevant and thus generically driven to larger values under the RG flow, with its IR value being a free parameter of the model.
In the left panel of Fig.~\ref{fig:thetasAndRunning}, we show how the critical exponents depend on $Y_{\nu}$ to explicitly show that $\theta_{\nu}<0$, and hence a vanishing neutrino mass, is caused by its vanishing charge $Y_{\nu}=0$. 

\begin{figure*}[ht]
\centering
\includegraphics[width=0.5\linewidth]{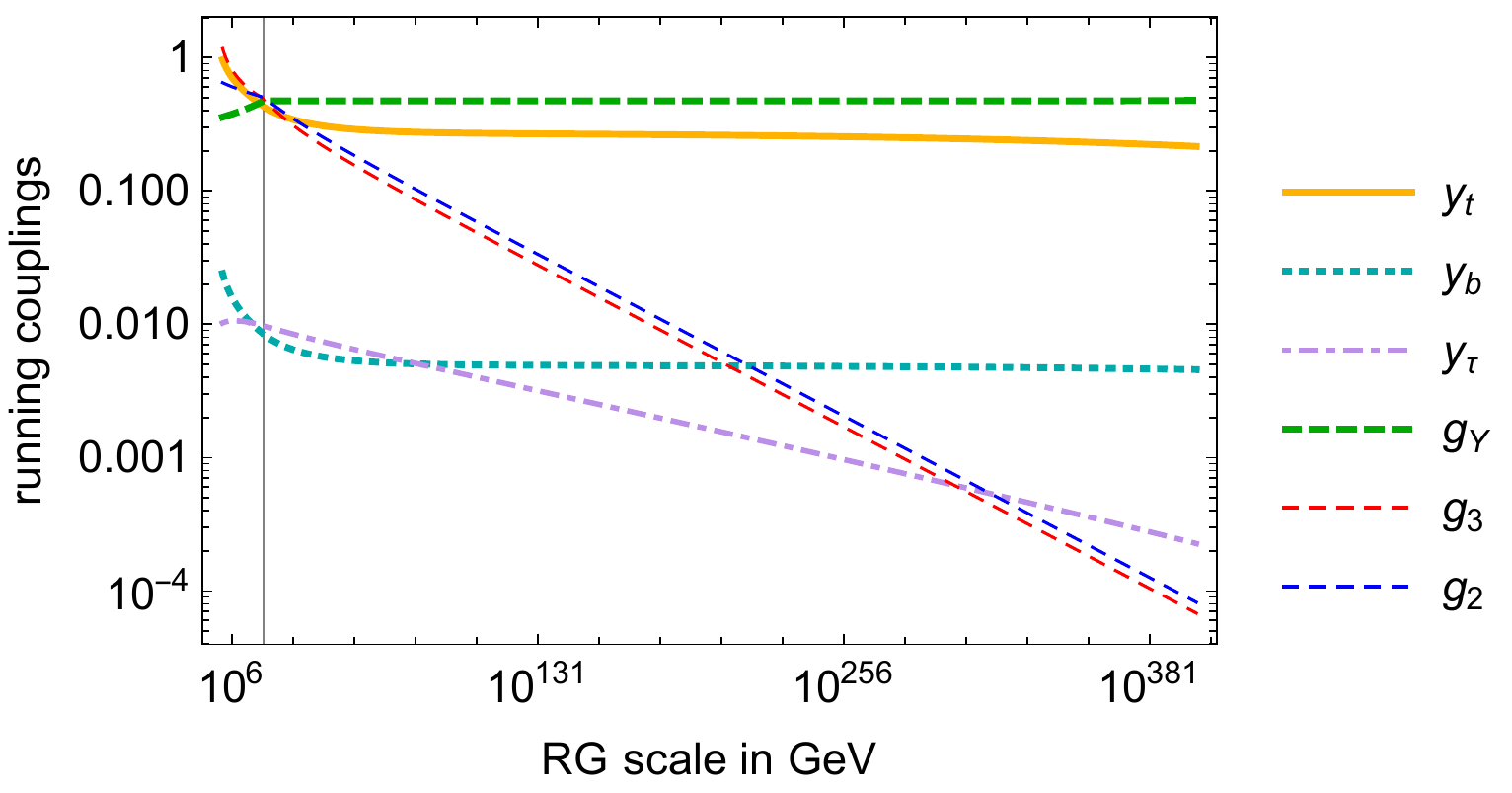} \quad \includegraphics[width=0.45\linewidth]{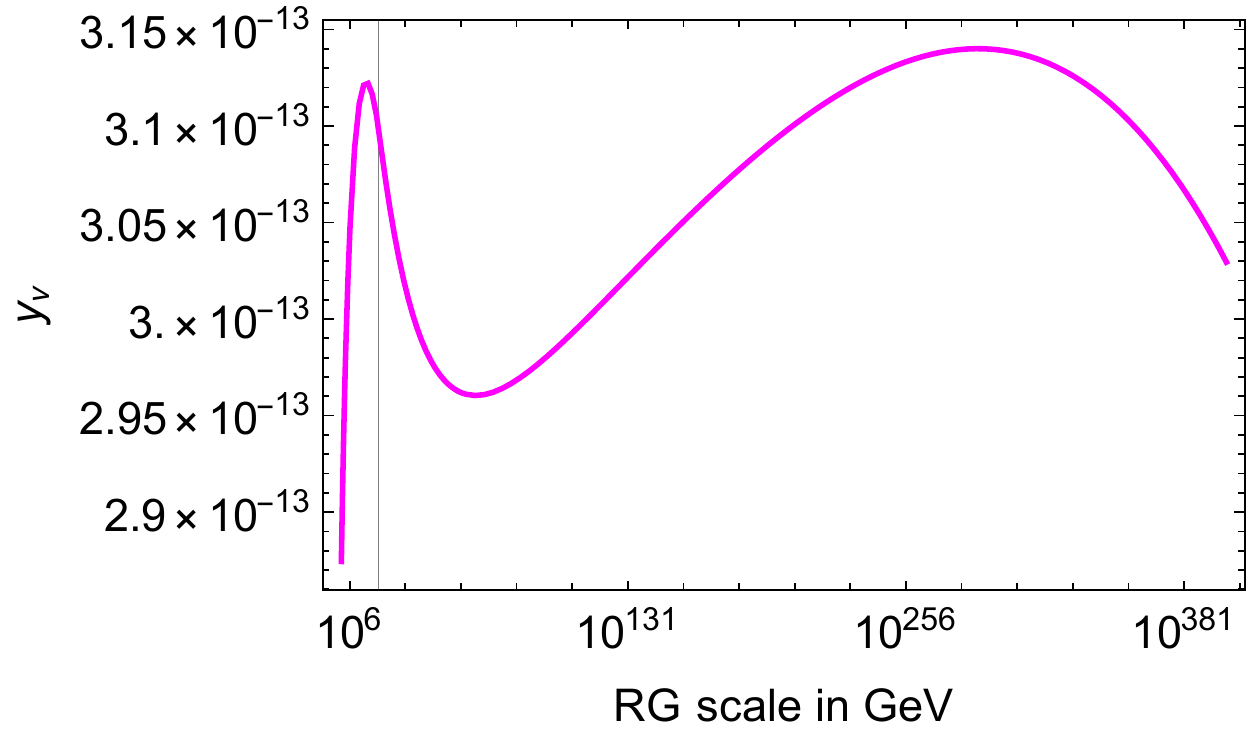}
\caption{\label{fig:tbyflow} 
We show the RG flow, starting from an asymptotically free (except for the Abelian gauge coupling, which is asymptotically safe) fixed point in the very deep UV. The deep UV regime is not shown, although most couplings already exhibit scaling towards the free fixed point in the range of scales shown here. All couplings reach IR values close to their SM values, but, due to an extended regime close to the interacting fixed point, the top Yukawa coupling is about 3~\% larger than its SM value. The neutrino-Yukawa (right panel) starts from an asymptotically free fixed point, exhibits a decrease when top- and bottom-Yukawa are close to their fixed-point values, and is finally drawn towards lower values in the sub-planckian regime. The corresponding neutrino mass is about 50 meV. We work with $f_g=9.7 \cdot 10^{-3}$ and $f_y= 1.188\cdot 10^{-4}$.}
\end{figure*}
\subsection{Absence of the maximally-predictive fixed point with four predictions for Yukawa couplings}

We a priori expect an interacting fixed point with four irrelevant directions which gives rise to a prediction for each of the four Yukawa couplings. This expectation is caused by the structure of the beta functions: schematically, $\beta_y = - f_y \, y + c_1 y+ c_2\, y^3$, with $c_1$ a contribution from the gauge couplings and other Yukawa couplings which is negative in the phenomenologically relevant heavy-top limit and $c_2$ a positive constant. From this structure and the choice $f_y>0$, an interacting fixed point $y_{\ast} =\sqrt{\frac{f_y+c_1}{c_2}}>0$ follows. At this fixed point, the critical exponent is negative, i.e., the fixed point generates a prediction for the IR value of the Yukawa coupling.\\
 Given that all four beta functions Eqs.~\eqref{eq:betayt3}-\eqref{eq:betayb3} and \eqref{eq:betaytau3}-\eqref{eq:betaynu3} have this structure, we expect a fixed point which is interacting in all four Yukawa couplings and generates four predictions in the IR. However, it turns out that the interacting-Yukawa fixed point identified in Eq.~\eqref{eq:interacting-FP}, which has three negative critical exponents and generates three predictions whenever it is physical, i.e., $f_y$ large enough for $y_{b,\,\ast}^2>0$, is one of the most predictive fixed points of the system. 
We show below that a fixed point with four predictions is absent, because SM quarks and leptons couple to the same Higgs doublet.  

The anomalous dimension of the Higgs is generated by fermionic loops of all fermions which couple to the same Higgs field. This anomalous dimension contributes to the $\beta$ function for all Yukawa couplings which couple a given fermion to the same Higgs field.
Because the quarks and leptons of the SM couple to a joint Higgs doublet, we obtain the terms $(y_\tau^2 + y_{\nu}^2)$ and $(3y_t^2 + 3y_b^2)$ in Eqs.~\eqref{eq:betayt3}-\eqref{eq:betayb3} and \eqref{eq:betaytau3}-\eqref{eq:betaynu3}, respectively. 

This causes the fully interacting fixed point to vanish. To make this more explicit, we introduce a fiducial parameter $\epsilon$ and replace $(y_\tau^2 + y_{\nu_\tau}^2)\rightarrow \epsilon(y_\tau^2 + y_{\nu_\tau}^2)$ in $\beta_{y_{t}}$ and $\beta_{y_{b}}$ as well as $(3y_t^2 + 3y_b^2)\rightarrow \epsilon(3y_t^2 + 3y_b^2)$ in $\beta_{y_{\tau}}$ and $\beta_{y_{\nu}}$. For $\epsilon\rightarrow0$, quarks and leptons couple to separate Higgs fields while for $\epsilon\rightarrow1$ one recovers the SM case in which they all couple to a joint Higgs field. 

For any $\epsilon\neq 1$ we find a fully interacting fixed point at which all Yukawas are proportional to $y_{t/b/\tau/\nu\,\ast}^2\sim 1/\left(\epsilon ^2-1\right)$. In the SM limit ($\epsilon\rightarrow1$) the fully interacting fixed point diverges, i.e., $y_{t/b/\tau/\nu\,\ast}\rightarrow\infty$. 
Therefore, the interacting-Yukawa fixed point in Eq.~\eqref{eq:interacting-FP} leads to the maximal possible number of predictions. 

\section{Prediction from a fixed point at finite top Yukawa coupling: vanishing Dirac neutrino mass}

The only asymptotically safe fixed point at non-zero top- and bottom-Yukawa coupling which is not phenomenologically excluded lies at $y_{\nu\, \ast}= 0 = y_{\tau\, \ast}$, cf.~Tab.~\ref{tab:fps}. From this fixed-point, $y_{\tau}$ can depart, because it corresponds to a relevant direction ($\theta_{\tau}>0$). In contrast, $\theta_{\nu}<0$, which implies that the neutrino Yukawa coupling must remain zero.

Therefore, the extension of the work in \cite{Eichhorn:2018whv} to the lepton sector is phenomenologically quite intriguing: those three Yukawa couplings of the third generation that have been measured to be nonzero at the LHC, namely top quark Yukawa \cite{CMS:2018uxb,ATLAS:2018mme}, bottom quark Yukawa \cite{CMS:2018nsn,ATLAS:2018kot} and tau lepton Yukawa \cite{ATLAS:2015xst,CMS:2017zyp}, are either predicted in the vicinity of the SM values or can be accommodated at exactly the SM value. Further, the fixed point results in a prediction for the single Yukawa coupling of the third generation which has not yet been measured, namely the neutrino Yukawa coupling. The prediction implies that the neutrino cannot have a Dirac mass term.

The prediction of the top-quark Yukawa coupling does not match the observed SM value exactly. This may well be a consequence of our simplified treatment which is limited to the one-loop SM terms and a parametrized quantum-gravity contribution. Higher-order effects (corresponding to effects of higher-order operators in an FRG treatment as well as FRG threshold effects, both at the Planck scale and the electroweak scale) are thus neglected. These may generate quantitative corrections under the flow. It is important to stress that such terms cannot change the conclusion that the neutrino Yukawa coupling vanishes at all scales. The latter conclusion relies on the fact that there is a separate global U(1) symmetry for each SM fermion, if its Yukawa coupling vanishes. 
Therefore, a Yukawa coupling, once set to zero, cannot be generated by radiative corrections. If the coupling is irrelevant, then it cannot depart from zero anywhere above the Planck scale. Below the Planck scale, the enhanced global symmetry prohibits radiative corrections that would generate the neutrino's Yukawa coupling.
\\

\section{Cross-over trajectories: UV completeness and meV-scale Dirac neutrino mass}
\begin{figure*}[t]
\centering
\includegraphics[width=0.5\linewidth]{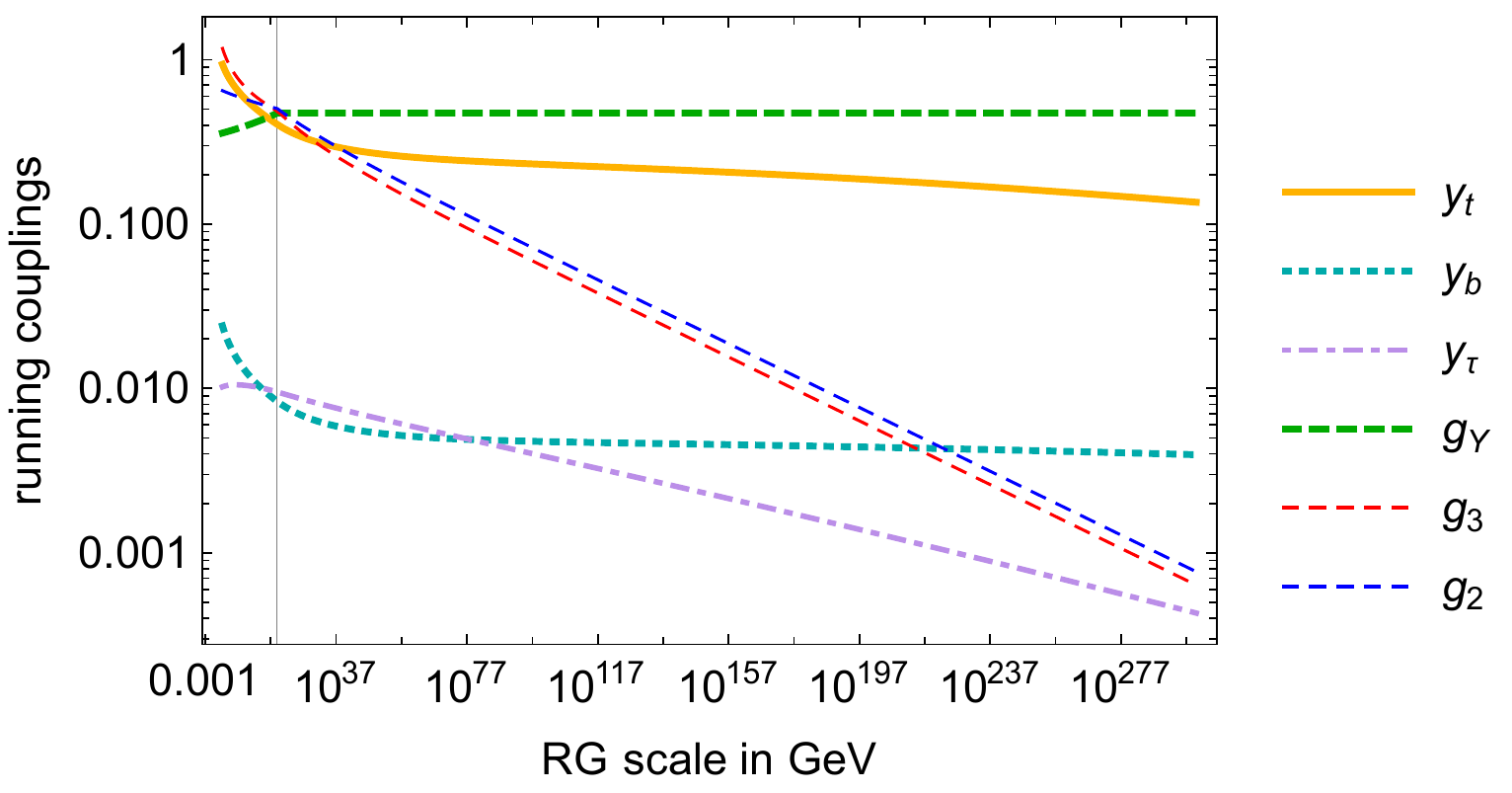} \quad \includegraphics[width=0.45\linewidth]{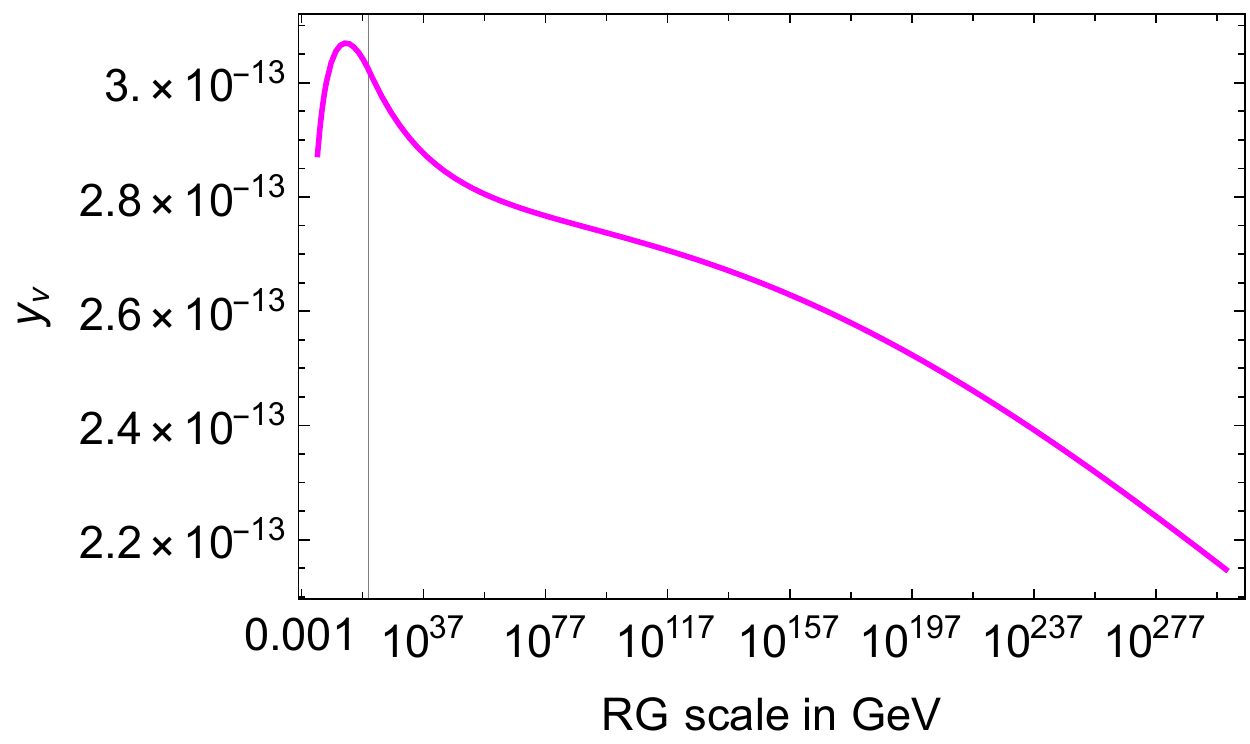}
\caption{\label{fig:SMflow} We show the RG flow, starting from an asymptotically free (except for the Abelian gauge coupling, which is asymptotically safe) fixed point in the very deep UV. The deep UV regime is not shown, although most couplings already exhibit scaling towards the free fixed point in the range of scales shown here. All couplings reach their SM values in the IR. The neutrino-Yukawa (right panel) starts from an asymptotically free fixed point, exhibits a slower running when the top- and bottom-Yukawa are close to the interacting fixed point, and is finally drawn towards lower values in the sub-planckian regime. The corresponding neutrino mass is about 50 meV. We work with $f_g=9.7 \cdot 10^{-3}$ and $f_y= 1.188\cdot 10^{-4}$. }
\end{figure*}

In the previous section, we saw that an asymptotically safe fixed point gives rise to a prediction of the top-, bottom and neutrino-Yukawa. It predicts top- and bottom mass in the vicinity of their SM values, and a vanishing Dirac neutrino mass. In this scenario, neutrinos can only be massive, if a mass-generation mechanism does not rely on a neutrino Yukawa coupling.
\\

Within the asymptotic-safety paradigm, there is also a  possibility for a finite Dirac neutrino mass, based on a cross-over trajectory from a fixed point which is non-interacting in the Yukawa couplings (the free-Yukawa fixed point) to a fixed point which is interacting in the Yukawa couplings (the safe-Yukawa fixed point), see Tab.~\ref{tab:fps}. (At both fixed points, $g_Y$ is interacting, thus both fixed points are asymptotically safe.)
\\
Under the RG flow in the transplanckian regime, $g_Y$ stays constant, while all other couplings increase from their vanishing fixed-point values. $y_t$ and $y_b$ are attracted to their interacting fixed-point values. Once they are close enough to their fixed-point values, $y_{\nu}$ can no longer continue to increase, because its critical exponent is negative, see Tab.~\ref{tab:fps}. Accordingly, $y_{\nu}$ is attracted towards $y_{\nu\, \ast}=0$ and shrinks, cf.~Fig.~\ref{fig:tbyflow}.\footnote{A similar behavior of $y_{\nu}$ is realized along a cross-over trajectory to another fixed point, not listed in Tab.~\ref{tab:fps}. At that fixed point, all Yukawas except the top quark Yukawa vanish, $y_{t\, \ast}>0$. The bottom Yukawa coupling is irrelevant, if $f_y$ is chosen such that $y_{t\, \rm IR}$ comes out correctly. This rules out the fixed point as a fundamental fixed point, but its vicinity can be reached along a cross-over trajectory. The neutrino Yukawa coupling is irrelevant at $y_{\nu\, \ast}=0$, thus this fixed point similarly generates a small neutrino mass.} 

The qualitative features of such cross-over trajectories are very intriguing, because the neutrino Yukawa and the tau Yukawa coupling behave quite distinctly: whereas the tau Yukawa coupling rises monotonically starting from the free fixed point, the neutrino Yukawa coupling slows its increase or even decreases. Thus, the neutrino Yukawa becomes naturally smaller than the tau Yukawa coupling along these trajectories.

The neutrino Yukawa shrinks the more, the longer the range of scales that top and bottom Yukawa spend in the fixed-point regime. This can be quantified by the critical exponent $\theta_{\nu} = - \frac{f_g}{41}$. Thus, in the fixed-point regime
\be
y_{\nu}(k) = y_{\nu}(k_{\rm UV}) \left(\frac{k}{k_{\rm UV}}\right)^{-\theta_\nu}
\ee
where $k_{\rm UV}$ is a far transplanckian scale at which the flow reaches the vicinity of the interacting fixed point. For the value $f_g=9.8\times10^{-4}$, it takes a ratio $k/k_{\rm UV} \approx 10^{-4183\, n}$ for the neutrino Yukawa to decrease by $n$ orders of magnitude, i.e., the small critical exponent of the neutrino translates into a slow flow. At a first glance, the range of scales required for a significant decrease of the neutrino Yukawa coupling may seem unusual, in the sense that considerations in beyond-SM settings are typically restricted to much smaller ranges of scales. However, these considerations are typically made in the context of effective theories with a finite UV cutoff (which is often the Planck scale), and thus of course the available range of scales is restricted. In contrast, the situation is quite different in our case. We explore a setting in which there is no UV limit to how far the theory may be valid. In the face of a literally infinite range of scales, comparatively slow flows which require large ranges of scales to produce a significant change in the values of couplings, are no reason for skepticism or surprise.

\section{Discussion and outlook}
\label{sec:discussion}

In this letter, we have found evidence that asymptotic safety of gravity and matter dynamically leads to a small Dirac neutrino mass.
The Dirac neutrino mass is small, because asymptotic safety favors a small neutrino Yukawa coupling. At the interacting-Yukawa fixed point, the neutrino Yukawa coupling is zero and irrelevant. In other words, quantum fluctuations of gravity and matter drive the neutrino Yukawa coupling towards zero. This property also impacts cross-over trajectories that start from the free-Yukawa fixed point and pass close to the interacting-Yukawa fixed point on their way to SM-values in the IR, cf.~Fig.~\ref{fig:SMflow}.

The irrelevance of the neutrino Yukawa coupling is linked directly to the vanishing hypercharge of the right-handed neutrino, cf.~Fig.~\ref{fig:thetasAndRunning}. If the right-handed neutrino was not sterile, but had a hypercharge of order 1, similar to the other leptons and quarks, the neutrino Yukawa coupling would be relevant at the interacting-Yukawa fixed point. Then, in turn, asymptotic safety would no longer favor small Dirac neutrino masses.\\
This is a second example of a mechanism in asymptotic safety which ties masses to charges: At the interacting-Yukawa fixed point, a mass-splitting between the top and bottom quark is tied to their different hypercharges \cite{Eichhorn:2018whv}; additionally, we find that a tiny Dirac neutrino mass is also tied to the hypercharge assignment.\\

Our results have phenomenological consequences for the nature and generation of neutrino masses. We discuss these, keeping in mind that we have obtained our result in a study of the third generation of SM fermions and some properties might change in an extension to all three generations. \footnote{During the final stages of this work, a closely related paper \cite{Kowalska:2022ypk} with similar results has appeared which confirms that key properties relevant for phenomenology do in fact extend to the case of three generations.}\\
First, our results suggest that asymptotic safety does not need a seesaw mechanism to obtain small neutrino masses. Not only can a small Dirac neutrino mass be accommodated, it is in fact \emph{favored} by the asymptotically safe fixed-point structure in our study. In particular, cross-over trajectories from the free-Yukawa fixed point exhibit a long regime of limited growth or even decrease of the neutrino Yukawa coupling.\\
Second, starting from the interacting-Yukawa fixed point\footnote{Whether or not this fixed point is compatible with SM phenomenology, is currently an open question. Within the setting without higher-order operators resulting in subleading corrections and Planck-scale threshold effects, the top quark mass is predicted to be in the vicinity, but not in agreement, with the SM result at 178 GeV. A similar result holds once three generations of quarks are considered \cite{Alkofer:2020vtb}.
If this property persists, once the systematic uncertainties of our study are reduced, then cross-over trajectories from the free-Yukawa fixed point are the phenomenologically relevant ones}, the Dirac neutrino mass vanishes exactly in the IR. Thus, seesaw mechanisms which combine Dirac and Majorana mass, are also ruled out. On interacting-Yukawa- trajectories, an alternative mechanism for mass generation is needed. Whether it may be tied to the Weinberg operator and closely related to radiative mass generation \cite{Cai:2017jrq,Klein:2019iws} is currently under investigation \cite{deBrito:2022}. \\

Our results call for extended studies which reduce various systematic uncertainties in order to obtain precise predictions from asymptotic safety that can be compared to experimental data.\\
First, $f_y$ and $f_g$ need to be calculated with higher precision, which mainly requires a systematic extension to higher-order operators in the gravitational sector.\\
Second, subleading corrections to $f_y$ need to be studied. These are linked to higher-order operators of the matter fields (e.g., four-fermion operators) and are expected to distinguish the different flavors and be largest for the top quark Yukawa coupling.\\
Third, an extension to the case of three generations, which goes beyond the parameterized approach in \cite{Alkofer:2020vtb,Kowalska:2022ypk} and accounts for the first two points, is required.\\
Fourth, threshold effects at the two transition scales (the Planck scale and the electroweak scale) need to be calculated within the FRG framework. Our current treatment is simplified in that we implement a hard transition at the Planck scale. Further, we assume that directly below the Planck scale, a standard perturbative treatment holds, with the well-investigated threshold/matching prescription at the electroweak scale. Describing both transitions within the FRG has the advantage that decoupling of massive modes occurs automatically and no matching prescriptions are needed.\\
Fifth, for a comprehensive treatment the Higgs potential needs to be included. Because our scenario relies on a fixed point at finite hypercharge gauge coupling, the Higgs potential at the Planck scale is not flat, as assumed in \cite{Shaposhnikov:2009pv}, but the Higgs quartic coupling remains irrelevant. Then, BSM physics, e.g., in the form of a Higgs portal to a dark matter sector, explored in asymptotic safety in \cite{Eichhorn:2017als,Reichert:2019car,Eichhorn:2020kca}, may be necessary to obtain the correct value of the Higgs mass \cite{Eichhorn:2021tsx}.

In summary, asymptotic safety may provide an attractive and highly predictive paradigm to answer open questions in and beyond the SM, such as, e.g., the smallness of neutrino masses. Due to its predictive power, it may also be ruled out based on experimental data from particle physics. In this property, it stands out from a large group of quantum-gravity approaches and gives meaning to the idea of experimental tests of quantum gravity.

\acknowledgements
We thank Reinhard Alkofer, Carlos Nieto, Roberto Percacci and Markus Schr\"ofl for discussions. A.~E.~is supported by a research grant (29405) from VILLUM fonden. 
The work leading to this publication was supported by the PRIME programme of the German Academic Exchange Service (DAAD) with funds from the German Federal Ministry of Education and Research (BMBF).

\newpage
\bibliography{References}

\begin{thebibliography}{128}%
\makeatletter
\providecommand \@ifxundefined [1]{%
 \@ifx{#1\undefined}
}%
\providecommand \@ifnum [1]{%
 \ifnum #1\expandafter \@firstoftwo
 \else \expandafter \@secondoftwo
 \fi
}%
\providecommand \@ifx [1]{%
 \ifx #1\expandafter \@firstoftwo
 \else \expandafter \@secondoftwo
 \fi
}%
\providecommand \natexlab [1]{#1}%
\providecommand \enquote  [1]{``#1''}%
\providecommand \bibnamefont  [1]{#1}%
\providecommand \bibfnamefont [1]{#1}%
\providecommand \citenamefont [1]{#1}%
\providecommand \href@noop [0]{\@secondoftwo}%
\providecommand \href [0]{\begingroup \@sanitize@url \@href}%
\providecommand \@href[1]{\@@startlink{#1}\@@href}%
\providecommand \@@href[1]{\endgroup#1\@@endlink}%
\providecommand \@sanitize@url [0]{\catcode `\\12\catcode `\$12\catcode
  `\&12\catcode `\#12\catcode `\^12\catcode `\_12\catcode `\%12\relax}%
\providecommand \@@startlink[1]{}%
\providecommand \@@endlink[0]{}%
\providecommand \url  [0]{\begingroup\@sanitize@url \@url }%
\providecommand \@url [1]{\endgroup\@href {#1}{\urlprefix }}%
\providecommand \urlprefix  [0]{URL }%
\providecommand \Eprint [0]{\href }%
\providecommand \doibase [0]{http://dx.doi.org/}%
\providecommand \selectlanguage [0]{\@gobble}%
\providecommand \bibinfo  [0]{\@secondoftwo}%
\providecommand \bibfield  [0]{\@secondoftwo}%
\providecommand \translation [1]{[#1]}%
\providecommand \BibitemOpen [0]{}%
\providecommand \bibitemStop [0]{}%
\providecommand \bibitemNoStop [0]{.\EOS\space}%
\providecommand \EOS [0]{\spacefactor3000\relax}%
\providecommand \BibitemShut  [1]{\csname bibitem#1\endcsname}%
\let\auto@bib@innerbib\@empty
\bibitem [{\citenamefont {Zyla}\ \emph {et~al.}(2020)\citenamefont {Zyla} \emph
  {et~al.}}]{ParticleDataGroup:2020ssz}%
  \BibitemOpen
  \bibfield  {author} {\bibinfo {author} {\bibfnamefont {P.~A.}\ \bibnamefont
  {Zyla}} \emph {et~al.} (\bibinfo {collaboration} {Particle Data Group}),\
  }\href {\doibase 10.1093/ptep/ptaa104} {\bibfield  {journal} {\bibinfo
  {journal} {PTEP}\ }\textbf {\bibinfo {volume} {2020}},\ \bibinfo {pages}
  {083C01} (\bibinfo {year} {2020})}\BibitemShut {NoStop}%
\bibitem [{\citenamefont {Davis}\ \emph {et~al.}(1968)\citenamefont {Davis},
  \citenamefont {Harmer},\ and\ \citenamefont {Hoffman}}]{Davis:1968cp}%
  \BibitemOpen
  \bibfield  {author} {\bibinfo {author} {\bibfnamefont {R.}~\bibnamefont
  {Davis}, \bibfnamefont {Jr.}}, \bibinfo {author} {\bibfnamefont {D.~S.}\
  \bibnamefont {Harmer}}, \ and\ \bibinfo {author} {\bibfnamefont {K.~C.}\
  \bibnamefont {Hoffman}},\ }\href {\doibase 10.1103/PhysRevLett.20.1205}
  {\bibfield  {journal} {\bibinfo  {journal} {Phys. Rev. Lett.}\ }\textbf
  {\bibinfo {volume} {20}},\ \bibinfo {pages} {1205} (\bibinfo {year}
  {1968})}\BibitemShut {NoStop}%
\bibitem [{\citenamefont {Cleveland}\ \emph {et~al.}(1998)\citenamefont
  {Cleveland}, \citenamefont {Daily}, \citenamefont {Davis}, \citenamefont
  {Distel}, \citenamefont {Lande}, \citenamefont {Lee}, \citenamefont
  {Wildenhain},\ and\ \citenamefont {Ullman}}]{Cleveland:1998nv}%
  \BibitemOpen
  \bibfield  {author} {\bibinfo {author} {\bibfnamefont {B.~T.}\ \bibnamefont
  {Cleveland}}, \bibinfo {author} {\bibfnamefont {T.}~\bibnamefont {Daily}},
  \bibinfo {author} {\bibfnamefont {R.}~\bibnamefont {Davis}, \bibfnamefont
  {Jr.}}, \bibinfo {author} {\bibfnamefont {J.~R.}\ \bibnamefont {Distel}},
  \bibinfo {author} {\bibfnamefont {K.}~\bibnamefont {Lande}}, \bibinfo
  {author} {\bibfnamefont {C.~K.}\ \bibnamefont {Lee}}, \bibinfo {author}
  {\bibfnamefont {P.~S.}\ \bibnamefont {Wildenhain}}, \ and\ \bibinfo {author}
  {\bibfnamefont {J.}~\bibnamefont {Ullman}},\ }\href {\doibase 10.1086/305343}
  {\bibfield  {journal} {\bibinfo  {journal} {Astrophys. J.}\ }\textbf
  {\bibinfo {volume} {496}},\ \bibinfo {pages} {505} (\bibinfo {year}
  {1998})}\BibitemShut {NoStop}%
\bibitem [{\citenamefont {Fukuda}\ \emph {et~al.}(1996)\citenamefont {Fukuda}
  \emph {et~al.}}]{Fukuda:1996sz}%
  \BibitemOpen
  \bibfield  {author} {\bibinfo {author} {\bibfnamefont {Y.}~\bibnamefont
  {Fukuda}} \emph {et~al.} (\bibinfo {collaboration} {Kamiokande}),\ }\href
  {\doibase 10.1103/PhysRevLett.77.1683} {\bibfield  {journal} {\bibinfo
  {journal} {Phys. Rev. Lett.}\ }\textbf {\bibinfo {volume} {77}},\ \bibinfo
  {pages} {1683} (\bibinfo {year} {1996})}\BibitemShut {NoStop}%
\bibitem [{\citenamefont {Hampel}\ \emph {et~al.}(1999)\citenamefont {Hampel}
  \emph {et~al.}}]{Hampel:1998xg}%
  \BibitemOpen
  \bibfield  {author} {\bibinfo {author} {\bibfnamefont {W.}~\bibnamefont
  {Hampel}} \emph {et~al.} (\bibinfo {collaboration} {GALLEX}),\ }\href
  {\doibase 10.1016/S0370-2693(98)01579-2} {\bibfield  {journal} {\bibinfo
  {journal} {Phys. Lett. B}\ }\textbf {\bibinfo {volume} {447}},\ \bibinfo
  {pages} {127} (\bibinfo {year} {1999})}\BibitemShut {NoStop}%
\bibitem [{\citenamefont {Fukuda}\ \emph {et~al.}(2002)\citenamefont {Fukuda}
  \emph {et~al.}}]{Fukuda:2002pe}%
  \BibitemOpen
  \bibfield  {author} {\bibinfo {author} {\bibfnamefont {S.}~\bibnamefont
  {Fukuda}} \emph {et~al.} (\bibinfo {collaboration} {Super-Kamiokande}),\
  }\href {\doibase 10.1016/S0370-2693(02)02090-7} {\bibfield  {journal}
  {\bibinfo  {journal} {Phys. Lett. B}\ }\textbf {\bibinfo {volume} {539}},\
  \bibinfo {pages} {179} (\bibinfo {year} {2002})},\ \Eprint
  {http://arxiv.org/abs/hep-ex/0205075} {arXiv:hep-ex/0205075} \BibitemShut
  {NoStop}%
\bibitem [{\citenamefont {Ahmad}\ \emph {et~al.}(2002)\citenamefont {Ahmad}
  \emph {et~al.}}]{Ahmad:2002jz}%
  \BibitemOpen
  \bibfield  {author} {\bibinfo {author} {\bibfnamefont {Q.~R.}\ \bibnamefont
  {Ahmad}} \emph {et~al.} (\bibinfo {collaboration} {SNO}),\ }\href {\doibase
  10.1103/PhysRevLett.89.011301} {\bibfield  {journal} {\bibinfo  {journal}
  {Phys. Rev. Lett.}\ }\textbf {\bibinfo {volume} {89}},\ \bibinfo {pages}
  {011301} (\bibinfo {year} {2002})},\ \Eprint
  {http://arxiv.org/abs/nucl-ex/0204008} {arXiv:nucl-ex/0204008} \BibitemShut
  {NoStop}%
\bibitem [{\citenamefont {Altmann}\ \emph {et~al.}(2005)\citenamefont {Altmann}
  \emph {et~al.}}]{Altmann:2005ix}%
  \BibitemOpen
  \bibfield  {author} {\bibinfo {author} {\bibfnamefont {M.}~\bibnamefont
  {Altmann}} \emph {et~al.} (\bibinfo {collaboration} {GNO}),\ }\href {\doibase
  10.1016/j.physletb.2005.04.068} {\bibfield  {journal} {\bibinfo  {journal}
  {Phys. Lett. B}\ }\textbf {\bibinfo {volume} {616}},\ \bibinfo {pages} {174}
  (\bibinfo {year} {2005})},\ \Eprint {http://arxiv.org/abs/hep-ex/0504037}
  {arXiv:hep-ex/0504037} \BibitemShut {NoStop}%
\bibitem [{\citenamefont {Abdurashitov}\ \emph {et~al.}(2009)\citenamefont
  {Abdurashitov} \emph {et~al.}}]{Abdurashitov:2009tn}%
  \BibitemOpen
  \bibfield  {author} {\bibinfo {author} {\bibfnamefont {J.~N.}\ \bibnamefont
  {Abdurashitov}} \emph {et~al.} (\bibinfo {collaboration} {SAGE}),\ }\href
  {\doibase 10.1103/PhysRevC.80.015807} {\bibfield  {journal} {\bibinfo
  {journal} {Phys. Rev. C}\ }\textbf {\bibinfo {volume} {80}},\ \bibinfo
  {pages} {015807} (\bibinfo {year} {2009})},\ \Eprint
  {http://arxiv.org/abs/0901.2200} {arXiv:0901.2200 [nucl-ex]} \BibitemShut
  {NoStop}%
\bibitem [{\citenamefont {Hirata}\ \emph {et~al.}(1988)\citenamefont {Hirata}
  \emph {et~al.}}]{Kamiokande-II:1988sxn}%
  \BibitemOpen
  \bibfield  {author} {\bibinfo {author} {\bibfnamefont {K.~S.}\ \bibnamefont
  {Hirata}} \emph {et~al.} (\bibinfo {collaboration} {Kamiokande-II}),\ }\href
  {\doibase 10.1016/0370-2693(88)91690-5} {\bibfield  {journal} {\bibinfo
  {journal} {Phys. Lett. B}\ }\textbf {\bibinfo {volume} {205}},\ \bibinfo
  {pages} {416} (\bibinfo {year} {1988})}\BibitemShut {NoStop}%
\bibitem [{\citenamefont {Casper}\ \emph {et~al.}(1991)\citenamefont {Casper}
  \emph {et~al.}}]{Casper:1990ac}%
  \BibitemOpen
  \bibfield  {author} {\bibinfo {author} {\bibfnamefont {D.}~\bibnamefont
  {Casper}} \emph {et~al.},\ }\href {\doibase 10.1103/PhysRevLett.66.2561}
  {\bibfield  {journal} {\bibinfo  {journal} {Phys. Rev. Lett.}\ }\textbf
  {\bibinfo {volume} {66}},\ \bibinfo {pages} {2561} (\bibinfo {year}
  {1991})}\BibitemShut {NoStop}%
\bibitem [{\citenamefont {Fukuda}\ \emph {et~al.}(1998)\citenamefont {Fukuda}
  \emph {et~al.}}]{Fukuda:1998mi}%
  \BibitemOpen
  \bibfield  {author} {\bibinfo {author} {\bibfnamefont {Y.}~\bibnamefont
  {Fukuda}} \emph {et~al.} (\bibinfo {collaboration} {Super-Kamiokande}),\
  }\href {\doibase 10.1103/PhysRevLett.81.1562} {\bibfield  {journal} {\bibinfo
   {journal} {Phys. Rev. Lett.}\ }\textbf {\bibinfo {volume} {81}},\ \bibinfo
  {pages} {1562} (\bibinfo {year} {1998})},\ \Eprint
  {http://arxiv.org/abs/hep-ex/9807003} {arXiv:hep-ex/9807003} \BibitemShut
  {NoStop}%
\bibitem [{\citenamefont {Ashie}\ \emph {et~al.}(2004)\citenamefont {Ashie}
  \emph {et~al.}}]{Ashie:2004mr}%
  \BibitemOpen
  \bibfield  {author} {\bibinfo {author} {\bibfnamefont {Y.}~\bibnamefont
  {Ashie}} \emph {et~al.} (\bibinfo {collaboration} {Super-Kamiokande}),\
  }\href {\doibase 10.1103/PhysRevLett.93.101801} {\bibfield  {journal}
  {\bibinfo  {journal} {Phys. Rev. Lett.}\ }\textbf {\bibinfo {volume} {93}},\
  \bibinfo {pages} {101801} (\bibinfo {year} {2004})},\ \Eprint
  {http://arxiv.org/abs/hep-ex/0404034} {arXiv:hep-ex/0404034} \BibitemShut
  {NoStop}%
\bibitem [{\citenamefont {Eguchi}\ \emph {et~al.}(2003)\citenamefont {Eguchi}
  \emph {et~al.}}]{Eguchi:2002dm}%
  \BibitemOpen
  \bibfield  {author} {\bibinfo {author} {\bibfnamefont {K.}~\bibnamefont
  {Eguchi}} \emph {et~al.} (\bibinfo {collaboration} {KamLAND}),\ }\href
  {\doibase 10.1103/PhysRevLett.90.021802} {\bibfield  {journal} {\bibinfo
  {journal} {Phys. Rev. Lett.}\ }\textbf {\bibinfo {volume} {90}},\ \bibinfo
  {pages} {021802} (\bibinfo {year} {2003})},\ \Eprint
  {http://arxiv.org/abs/hep-ex/0212021} {arXiv:hep-ex/0212021} \BibitemShut
  {NoStop}%
\bibitem [{\citenamefont {Araki}\ \emph {et~al.}(2005)\citenamefont {Araki}
  \emph {et~al.}}]{Araki:2004mb}%
  \BibitemOpen
  \bibfield  {author} {\bibinfo {author} {\bibfnamefont {T.}~\bibnamefont
  {Araki}} \emph {et~al.} (\bibinfo {collaboration} {KamLAND}),\ }\href
  {\doibase 10.1103/PhysRevLett.94.081801} {\bibfield  {journal} {\bibinfo
  {journal} {Phys. Rev. Lett.}\ }\textbf {\bibinfo {volume} {94}},\ \bibinfo
  {pages} {081801} (\bibinfo {year} {2005})},\ \Eprint
  {http://arxiv.org/abs/hep-ex/0406035} {arXiv:hep-ex/0406035} \BibitemShut
  {NoStop}%
\bibitem [{\citenamefont {Ahn}\ \emph {et~al.}(2006)\citenamefont {Ahn} \emph
  {et~al.}}]{Ahn:2006zza}%
  \BibitemOpen
  \bibfield  {author} {\bibinfo {author} {\bibfnamefont {M.~H.}\ \bibnamefont
  {Ahn}} \emph {et~al.} (\bibinfo {collaboration} {K2K}),\ }\href {\doibase
  10.1103/PhysRevD.74.072003} {\bibfield  {journal} {\bibinfo  {journal} {Phys.
  Rev. D}\ }\textbf {\bibinfo {volume} {74}},\ \bibinfo {pages} {072003}
  (\bibinfo {year} {2006})},\ \Eprint {http://arxiv.org/abs/hep-ex/0606032}
  {arXiv:hep-ex/0606032} \BibitemShut {NoStop}%
\bibitem [{\citenamefont {Michael}\ \emph {et~al.}(2006)\citenamefont {Michael}
  \emph {et~al.}}]{Michael:2006rx}%
  \BibitemOpen
  \bibfield  {author} {\bibinfo {author} {\bibfnamefont {D.~G.}\ \bibnamefont
  {Michael}} \emph {et~al.} (\bibinfo {collaboration} {MINOS}),\ }\href
  {\doibase 10.1103/PhysRevLett.97.191801} {\bibfield  {journal} {\bibinfo
  {journal} {Phys. Rev. Lett.}\ }\textbf {\bibinfo {volume} {97}},\ \bibinfo
  {pages} {191801} (\bibinfo {year} {2006})},\ \Eprint
  {http://arxiv.org/abs/hep-ex/0607088} {arXiv:hep-ex/0607088} \BibitemShut
  {NoStop}%
\bibitem [{\citenamefont {Gando}\ \emph {et~al.}(2016)\citenamefont {Gando}
  \emph {et~al.}}]{KamLAND-Zen:2016pfg}%
  \BibitemOpen
  \bibfield  {author} {\bibinfo {author} {\bibfnamefont {A.}~\bibnamefont
  {Gando}} \emph {et~al.} (\bibinfo {collaboration} {KamLAND-Zen}),\ }\href
  {\doibase 10.1103/PhysRevLett.117.082503} {\bibfield  {journal} {\bibinfo
  {journal} {Phys. Rev. Lett.}\ }\textbf {\bibinfo {volume} {117}},\ \bibinfo
  {pages} {082503} (\bibinfo {year} {2016})},\ \bibinfo {note} {[Addendum:
  Phys.Rev.Lett. 117, 109903 (2016)]},\ \Eprint
  {http://arxiv.org/abs/1605.02889} {arXiv:1605.02889 [hep-ex]} \BibitemShut
  {NoStop}%
\bibitem [{\citenamefont {Aker}\ \emph {et~al.}(2019)\citenamefont {Aker} \emph
  {et~al.}}]{KATRIN:2019yun}%
  \BibitemOpen
  \bibfield  {author} {\bibinfo {author} {\bibfnamefont {M.}~\bibnamefont
  {Aker}} \emph {et~al.} (\bibinfo {collaboration} {KATRIN}),\ }\href {\doibase
  10.1103/PhysRevLett.123.221802} {\bibfield  {journal} {\bibinfo  {journal}
  {Phys. Rev. Lett.}\ }\textbf {\bibinfo {volume} {123}},\ \bibinfo {pages}
  {221802} (\bibinfo {year} {2019})},\ \Eprint
  {http://arxiv.org/abs/1909.06048} {arXiv:1909.06048 [hep-ex]} \BibitemShut
  {NoStop}%
\bibitem [{\citenamefont {Gonzalez-Garcia}\ \emph {et~al.}(2012)\citenamefont
  {Gonzalez-Garcia}, \citenamefont {Maltoni}, \citenamefont {Salvado},\ and\
  \citenamefont {Schwetz}}]{Gonzalez-Garcia:2012hef}%
  \BibitemOpen
  \bibfield  {author} {\bibinfo {author} {\bibfnamefont {M.~C.}\ \bibnamefont
  {Gonzalez-Garcia}}, \bibinfo {author} {\bibfnamefont {M.}~\bibnamefont
  {Maltoni}}, \bibinfo {author} {\bibfnamefont {J.}~\bibnamefont {Salvado}}, \
  and\ \bibinfo {author} {\bibfnamefont {T.}~\bibnamefont {Schwetz}},\ }\href
  {\doibase 10.1007/JHEP12(2012)123} {\bibfield  {journal} {\bibinfo  {journal}
  {JHEP}\ }\textbf {\bibinfo {volume} {12}},\ \bibinfo {pages} {123} (\bibinfo
  {year} {2012})},\ \Eprint {http://arxiv.org/abs/1209.3023} {arXiv:1209.3023
  [hep-ph]} \BibitemShut {NoStop}%
\bibitem [{\citenamefont {Esteban}\ \emph {et~al.}(2020)\citenamefont
  {Esteban}, \citenamefont {Gonzalez-Garcia}, \citenamefont {Maltoni},
  \citenamefont {Schwetz},\ and\ \citenamefont {Zhou}}]{Esteban:2020cvm}%
  \BibitemOpen
  \bibfield  {author} {\bibinfo {author} {\bibfnamefont {I.}~\bibnamefont
  {Esteban}}, \bibinfo {author} {\bibfnamefont {M.~C.}\ \bibnamefont
  {Gonzalez-Garcia}}, \bibinfo {author} {\bibfnamefont {M.}~\bibnamefont
  {Maltoni}}, \bibinfo {author} {\bibfnamefont {T.}~\bibnamefont {Schwetz}}, \
  and\ \bibinfo {author} {\bibfnamefont {A.}~\bibnamefont {Zhou}},\ }\href
  {\doibase 10.1007/JHEP09(2020)178} {\bibfield  {journal} {\bibinfo  {journal}
  {JHEP}\ }\textbf {\bibinfo {volume} {09}},\ \bibinfo {pages} {178} (\bibinfo
  {year} {2020})},\ \Eprint {http://arxiv.org/abs/2007.14792} {arXiv:2007.14792
  [hep-ph]} \BibitemShut {NoStop}%
\bibitem [{\citenamefont {Palanque-Delabrouille}\ \emph
  {et~al.}(2015)\citenamefont {Palanque-Delabrouille} \emph
  {et~al.}}]{Palanque-Delabrouille:2015pga}%
  \BibitemOpen
  \bibfield  {author} {\bibinfo {author} {\bibfnamefont {N.}~\bibnamefont
  {Palanque-Delabrouille}} \emph {et~al.},\ }\href {\doibase
  10.1088/1475-7516/2015/11/011} {\bibfield  {journal} {\bibinfo  {journal}
  {JCAP}\ }\textbf {\bibinfo {volume} {11}},\ \bibinfo {pages} {011} (\bibinfo
  {year} {2015})},\ \Eprint {http://arxiv.org/abs/1506.05976} {arXiv:1506.05976
  [astro-ph.CO]} \BibitemShut {NoStop}%
\bibitem [{\citenamefont {Aghanim}\ \emph {et~al.}(2016)\citenamefont {Aghanim}
  \emph {et~al.}}]{Aghanim:2016yuo}%
  \BibitemOpen
  \bibfield  {author} {\bibinfo {author} {\bibfnamefont {N.}~\bibnamefont
  {Aghanim}} \emph {et~al.} (\bibinfo {collaboration} {Planck}),\ }\href
  {\doibase 10.1051/0004-6361/201628890} {\bibfield  {journal} {\bibinfo
  {journal} {Astron. Astrophys.}\ }\textbf {\bibinfo {volume} {596}},\ \bibinfo
  {pages} {A107} (\bibinfo {year} {2016})},\ \Eprint
  {http://arxiv.org/abs/1605.02985} {arXiv:1605.02985 [astro-ph.CO]}
  \BibitemShut {NoStop}%
\bibitem [{\citenamefont {Aghanim}\ \emph {et~al.}(2020)\citenamefont {Aghanim}
  \emph {et~al.}}]{Planck:2018vyg}%
  \BibitemOpen
  \bibfield  {author} {\bibinfo {author} {\bibfnamefont {N.}~\bibnamefont
  {Aghanim}} \emph {et~al.} (\bibinfo {collaboration} {Planck}),\ }\href
  {\doibase 10.1051/0004-6361/201833910} {\bibfield  {journal} {\bibinfo
  {journal} {Astron. Astrophys.}\ }\textbf {\bibinfo {volume} {641}},\ \bibinfo
  {pages} {A6} (\bibinfo {year} {2020})},\ \bibinfo {note} {[Erratum:
  Astron.Astrophys. 652, C4 (2021)]},\ \Eprint
  {http://arxiv.org/abs/1807.06209} {arXiv:1807.06209 [astro-ph.CO]}
  \BibitemShut {NoStop}%
\bibitem [{\citenamefont {Minkowski}(1977)}]{Minkowski:1977sc}%
  \BibitemOpen
  \bibfield  {author} {\bibinfo {author} {\bibfnamefont {P.}~\bibnamefont
  {Minkowski}},\ }\href {\doibase 10.1016/0370-2693(77)90435-X} {\bibfield
  {journal} {\bibinfo  {journal} {Phys. Lett.}\ }\textbf {\bibinfo {volume}
  {67B}},\ \bibinfo {pages} {421} (\bibinfo {year} {1977})}\BibitemShut
  {NoStop}%
\bibitem [{\citenamefont {Yanagida}(1979)}]{Yanagida:1979as}%
  \BibitemOpen
  \bibfield  {author} {\bibinfo {author} {\bibfnamefont {T.}~\bibnamefont
  {Yanagida}},\ }\href@noop {} {\bibfield  {journal} {\bibinfo  {journal}
  {Conf. Proc. C}\ }\textbf {\bibinfo {volume} {7902131}},\ \bibinfo {pages}
  {95} (\bibinfo {year} {1979})}\BibitemShut {NoStop}%
\bibitem [{\citenamefont {Glashow}(1980)}]{Glashow:1979nm}%
  \BibitemOpen
  \bibfield  {author} {\bibinfo {author} {\bibfnamefont {S.~L.}\ \bibnamefont
  {Glashow}},\ }\href {\doibase 10.1007/978-1-4684-7197-7_15} {\bibfield
  {journal} {\bibinfo  {journal} {NATO Sci. Ser. B}\ }\textbf {\bibinfo
  {volume} {61}},\ \bibinfo {pages} {687} (\bibinfo {year} {1980})}\BibitemShut
  {NoStop}%
\bibitem [{\citenamefont {Gell-Mann}\ \emph {et~al.}(1979)\citenamefont
  {Gell-Mann}, \citenamefont {Ramond},\ and\ \citenamefont
  {Slansky}}]{Gell-Mann:1979vob}%
  \BibitemOpen
  \bibfield  {author} {\bibinfo {author} {\bibfnamefont {M.}~\bibnamefont
  {Gell-Mann}}, \bibinfo {author} {\bibfnamefont {P.}~\bibnamefont {Ramond}}, \
  and\ \bibinfo {author} {\bibfnamefont {R.}~\bibnamefont {Slansky}},\
  }\href@noop {} {\bibfield  {journal} {\bibinfo  {journal} {Conf. Proc. C}\
  }\textbf {\bibinfo {volume} {790927}},\ \bibinfo {pages} {315} (\bibinfo
  {year} {1979})},\ \Eprint {http://arxiv.org/abs/1306.4669} {arXiv:1306.4669
  [hep-th]} \BibitemShut {NoStop}%
\bibitem [{\citenamefont {Mohapatra}\ and\ \citenamefont
  {Senjanovic}(1980)}]{Mohapatra:1979ia}%
  \BibitemOpen
  \bibfield  {author} {\bibinfo {author} {\bibfnamefont {R.~N.}\ \bibnamefont
  {Mohapatra}}\ and\ \bibinfo {author} {\bibfnamefont {G.}~\bibnamefont
  {Senjanovic}},\ }\href {\doibase 10.1103/PhysRevLett.44.912} {\bibfield
  {journal} {\bibinfo  {journal} {Phys. Rev. Lett.}\ }\textbf {\bibinfo
  {volume} {44}},\ \bibinfo {pages} {912} (\bibinfo {year} {1980})}\BibitemShut
  {NoStop}%
\bibitem [{\citenamefont {Magg}\ and\ \citenamefont
  {Wetterich}(1980)}]{Magg:1980ut}%
  \BibitemOpen
  \bibfield  {author} {\bibinfo {author} {\bibfnamefont {M.}~\bibnamefont
  {Magg}}\ and\ \bibinfo {author} {\bibfnamefont {C.}~\bibnamefont
  {Wetterich}},\ }\href {\doibase 10.1016/0370-2693(80)90825-4} {\bibfield
  {journal} {\bibinfo  {journal} {Phys. Lett. B}\ }\textbf {\bibinfo {volume}
  {94}},\ \bibinfo {pages} {61} (\bibinfo {year} {1980})}\BibitemShut {NoStop}%
\bibitem [{\citenamefont {Lazarides}\ \emph {et~al.}(1981)\citenamefont
  {Lazarides}, \citenamefont {Shafi},\ and\ \citenamefont
  {Wetterich}}]{Lazarides:1980nt}%
  \BibitemOpen
  \bibfield  {author} {\bibinfo {author} {\bibfnamefont {G.}~\bibnamefont
  {Lazarides}}, \bibinfo {author} {\bibfnamefont {Q.}~\bibnamefont {Shafi}}, \
  and\ \bibinfo {author} {\bibfnamefont {C.}~\bibnamefont {Wetterich}},\ }\href
  {\doibase 10.1016/0550-3213(81)90354-0} {\bibfield  {journal} {\bibinfo
  {journal} {Nucl. Phys. B}\ }\textbf {\bibinfo {volume} {181}},\ \bibinfo
  {pages} {287} (\bibinfo {year} {1981})}\BibitemShut {NoStop}%
\bibitem [{\citenamefont {Mohapatra}\ and\ \citenamefont
  {Senjanovic}(1981)}]{Mohapatra:1980yp}%
  \BibitemOpen
  \bibfield  {author} {\bibinfo {author} {\bibfnamefont {R.~N.}\ \bibnamefont
  {Mohapatra}}\ and\ \bibinfo {author} {\bibfnamefont {G.}~\bibnamefont
  {Senjanovic}},\ }\href {\doibase 10.1103/PhysRevD.23.165} {\bibfield
  {journal} {\bibinfo  {journal} {Phys. Rev. D}\ }\textbf {\bibinfo {volume}
  {23}},\ \bibinfo {pages} {165} (\bibinfo {year} {1981})}\BibitemShut
  {NoStop}%
\bibitem [{\citenamefont {Foot}\ \emph {et~al.}(1989)\citenamefont {Foot},
  \citenamefont {Lew}, \citenamefont {He},\ and\ \citenamefont
  {Joshi}}]{Foot:1988aq}%
  \BibitemOpen
  \bibfield  {author} {\bibinfo {author} {\bibfnamefont {R.}~\bibnamefont
  {Foot}}, \bibinfo {author} {\bibfnamefont {H.}~\bibnamefont {Lew}}, \bibinfo
  {author} {\bibfnamefont {X.~G.}\ \bibnamefont {He}}, \ and\ \bibinfo {author}
  {\bibfnamefont {G.~C.}\ \bibnamefont {Joshi}},\ }\href {\doibase
  10.1007/BF01415558} {\bibfield  {journal} {\bibinfo  {journal} {Z. Phys. C}\
  }\textbf {\bibinfo {volume} {44}},\ \bibinfo {pages} {441} (\bibinfo {year}
  {1989})}\BibitemShut {NoStop}%
\bibitem [{\citenamefont {King}(2004)}]{King:2003jb}%
  \BibitemOpen
  \bibfield  {author} {\bibinfo {author} {\bibfnamefont {S.~F.}\ \bibnamefont
  {King}},\ }\href {\doibase 10.1088/0034-4885/67/2/R01} {\bibfield  {journal}
  {\bibinfo  {journal} {Rept. Prog. Phys.}\ }\textbf {\bibinfo {volume} {67}},\
  \bibinfo {pages} {107} (\bibinfo {year} {2004})},\ \Eprint
  {http://arxiv.org/abs/hep-ph/0310204} {arXiv:hep-ph/0310204} \BibitemShut
  {NoStop}%
\bibitem [{\citenamefont {Mohapatra}\ and\ \citenamefont
  {Smirnov}(2006)}]{Mohapatra:2006gs}%
  \BibitemOpen
  \bibfield  {author} {\bibinfo {author} {\bibfnamefont {R.~N.}\ \bibnamefont
  {Mohapatra}}\ and\ \bibinfo {author} {\bibfnamefont {A.~Y.}\ \bibnamefont
  {Smirnov}},\ }\href {\doibase 10.1146/annurev.nucl.56.080805.140534}
  {\bibfield  {journal} {\bibinfo  {journal} {Ann. Rev. Nucl. Part. Sci.}\
  }\textbf {\bibinfo {volume} {56}},\ \bibinfo {pages} {569} (\bibinfo {year}
  {2006})},\ \Eprint {http://arxiv.org/abs/hep-ph/0603118}
  {arXiv:hep-ph/0603118} \BibitemShut {NoStop}%
\bibitem [{\citenamefont {Xing}\ and\ \citenamefont
  {Zhao}(2021)}]{Xing:2020ald}%
  \BibitemOpen
  \bibfield  {author} {\bibinfo {author} {\bibfnamefont {Z.-z.}\ \bibnamefont
  {Xing}}\ and\ \bibinfo {author} {\bibfnamefont {Z.-h.}\ \bibnamefont
  {Zhao}},\ }\href {\doibase 10.1088/1361-6633/abf086} {\bibfield  {journal}
  {\bibinfo  {journal} {Rept. Prog. Phys.}\ }\textbf {\bibinfo {volume} {84}},\
  \bibinfo {pages} {066201} (\bibinfo {year} {2021})},\ \Eprint
  {http://arxiv.org/abs/2008.12090} {arXiv:2008.12090 [hep-ph]} \BibitemShut
  {NoStop}%
\bibitem [{\citenamefont {Eichhorn}(2019)}]{Eichhorn:2018yfc}%
  \BibitemOpen
  \bibfield  {author} {\bibinfo {author} {\bibfnamefont {A.}~\bibnamefont
  {Eichhorn}},\ }\href {\doibase 10.3389/fspas.2018.00047} {\bibfield
  {journal} {\bibinfo  {journal} {Front. Astron. Space Sci.}\ }\textbf
  {\bibinfo {volume} {5}},\ \bibinfo {pages} {47} (\bibinfo {year} {2019})},\
  \Eprint {http://arxiv.org/abs/1810.07615} {arXiv:1810.07615 [hep-th]}
  \BibitemShut {NoStop}%
\bibitem [{\citenamefont {Eichhorn}(2022)}]{Eichhorn:2022jqj}%
  \BibitemOpen
  \bibfield  {author} {\bibinfo {author} {\bibfnamefont {A.}~\bibnamefont
  {Eichhorn}}\ }(\bibinfo {year} {2022})\ \Eprint
  {http://arxiv.org/abs/2201.11543} {arXiv:2201.11543 [gr-qc]} \BibitemShut
  {NoStop}%
\bibitem [{\citenamefont {Shaposhnikov}\ and\ \citenamefont
  {Wetterich}(2010)}]{Shaposhnikov:2009pv}%
  \BibitemOpen
  \bibfield  {author} {\bibinfo {author} {\bibfnamefont {M.}~\bibnamefont
  {Shaposhnikov}}\ and\ \bibinfo {author} {\bibfnamefont {C.}~\bibnamefont
  {Wetterich}},\ }\href {\doibase 10.1016/j.physletb.2009.12.022} {\bibfield
  {journal} {\bibinfo  {journal} {Phys. Lett. B}\ }\textbf {\bibinfo {volume}
  {683}},\ \bibinfo {pages} {196} (\bibinfo {year} {2010})},\ \Eprint
  {http://arxiv.org/abs/0912.0208} {arXiv:0912.0208 [hep-th]} \BibitemShut
  {NoStop}%
\bibitem [{\citenamefont {Harst}\ and\ \citenamefont
  {Reuter}(2011)}]{Harst:2011zx}%
  \BibitemOpen
  \bibfield  {author} {\bibinfo {author} {\bibfnamefont {U.}~\bibnamefont
  {Harst}}\ and\ \bibinfo {author} {\bibfnamefont {M.}~\bibnamefont {Reuter}},\
  }\href {\doibase 10.1007/JHEP05(2011)119} {\bibfield  {journal} {\bibinfo
  {journal} {JHEP}\ }\textbf {\bibinfo {volume} {05}},\ \bibinfo {pages} {119}
  (\bibinfo {year} {2011})},\ \Eprint {http://arxiv.org/abs/1101.6007}
  {arXiv:1101.6007 [hep-th]} \BibitemShut {NoStop}%
\bibitem [{\citenamefont {Eichhorn}\ and\ \citenamefont
  {Held}(2018{\natexlab{a}})}]{Eichhorn:2017ylw}%
  \BibitemOpen
  \bibfield  {author} {\bibinfo {author} {\bibfnamefont {A.}~\bibnamefont
  {Eichhorn}}\ and\ \bibinfo {author} {\bibfnamefont {A.}~\bibnamefont
  {Held}},\ }\href {\doibase 10.1016/j.physletb.2017.12.040} {\bibfield
  {journal} {\bibinfo  {journal} {Phys. Lett. B}\ }\textbf {\bibinfo {volume}
  {777}},\ \bibinfo {pages} {217} (\bibinfo {year} {2018}{\natexlab{a}})},\
  \Eprint {http://arxiv.org/abs/1707.01107} {arXiv:1707.01107 [hep-th]}
  \BibitemShut {NoStop}%
\bibitem [{\citenamefont {Eichhorn}\ and\ \citenamefont
  {Versteegen}(2018)}]{Eichhorn:2017lry}%
  \BibitemOpen
  \bibfield  {author} {\bibinfo {author} {\bibfnamefont {A.}~\bibnamefont
  {Eichhorn}}\ and\ \bibinfo {author} {\bibfnamefont {F.}~\bibnamefont
  {Versteegen}},\ }\href {\doibase 10.1007/JHEP01(2018)030} {\bibfield
  {journal} {\bibinfo  {journal} {JHEP}\ }\textbf {\bibinfo {volume} {01}},\
  \bibinfo {pages} {030} (\bibinfo {year} {2018})},\ \Eprint
  {http://arxiv.org/abs/1709.07252} {arXiv:1709.07252 [hep-th]} \BibitemShut
  {NoStop}%
\bibitem [{\citenamefont {Eichhorn}\ and\ \citenamefont
  {Held}(2018{\natexlab{b}})}]{Eichhorn:2018whv}%
  \BibitemOpen
  \bibfield  {author} {\bibinfo {author} {\bibfnamefont {A.}~\bibnamefont
  {Eichhorn}}\ and\ \bibinfo {author} {\bibfnamefont {A.}~\bibnamefont
  {Held}},\ }\href {\doibase 10.1103/PhysRevLett.121.151302} {\bibfield
  {journal} {\bibinfo  {journal} {Phys. Rev. Lett.}\ }\textbf {\bibinfo
  {volume} {121}},\ \bibinfo {pages} {151302} (\bibinfo {year}
  {2018}{\natexlab{b}})},\ \Eprint {http://arxiv.org/abs/1803.04027}
  {arXiv:1803.04027 [hep-th]} \BibitemShut {NoStop}%
\bibitem [{\citenamefont {Eichhorn}\ \emph
  {et~al.}(2018{\natexlab{a}})\citenamefont {Eichhorn}, \citenamefont {Held},\
  and\ \citenamefont {Wetterich}}]{Eichhorn:2017muy}%
  \BibitemOpen
  \bibfield  {author} {\bibinfo {author} {\bibfnamefont {A.}~\bibnamefont
  {Eichhorn}}, \bibinfo {author} {\bibfnamefont {A.}~\bibnamefont {Held}}, \
  and\ \bibinfo {author} {\bibfnamefont {C.}~\bibnamefont {Wetterich}},\ }\href
  {\doibase 10.1016/j.physletb.2018.05.016} {\bibfield  {journal} {\bibinfo
  {journal} {Phys. Lett. B}\ }\textbf {\bibinfo {volume} {782}},\ \bibinfo
  {pages} {198} (\bibinfo {year} {2018}{\natexlab{a}})},\ \Eprint
  {http://arxiv.org/abs/1711.02949} {arXiv:1711.02949 [hep-th]} \BibitemShut
  {NoStop}%
\bibitem [{\citenamefont {Eichhorn}\ \emph
  {et~al.}(2018{\natexlab{b}})\citenamefont {Eichhorn}, \citenamefont {Hamada},
  \citenamefont {Lumma},\ and\ \citenamefont {Yamada}}]{Eichhorn:2017als}%
  \BibitemOpen
  \bibfield  {author} {\bibinfo {author} {\bibfnamefont {A.}~\bibnamefont
  {Eichhorn}}, \bibinfo {author} {\bibfnamefont {Y.}~\bibnamefont {Hamada}},
  \bibinfo {author} {\bibfnamefont {J.}~\bibnamefont {Lumma}}, \ and\ \bibinfo
  {author} {\bibfnamefont {M.}~\bibnamefont {Yamada}},\ }\href {\doibase
  10.1103/PhysRevD.97.086004} {\bibfield  {journal} {\bibinfo  {journal} {Phys.
  Rev. D}\ }\textbf {\bibinfo {volume} {97}},\ \bibinfo {pages} {086004}
  (\bibinfo {year} {2018}{\natexlab{b}})},\ \Eprint
  {http://arxiv.org/abs/1712.00319} {arXiv:1712.00319 [hep-th]} \BibitemShut
  {NoStop}%
\bibitem [{\citenamefont {Eichhorn}\ \emph {et~al.}(2020)\citenamefont
  {Eichhorn}, \citenamefont {Held},\ and\ \citenamefont
  {Wetterich}}]{Eichhorn:2019dhg}%
  \BibitemOpen
  \bibfield  {author} {\bibinfo {author} {\bibfnamefont {A.}~\bibnamefont
  {Eichhorn}}, \bibinfo {author} {\bibfnamefont {A.}~\bibnamefont {Held}}, \
  and\ \bibinfo {author} {\bibfnamefont {C.}~\bibnamefont {Wetterich}},\ }\href
  {\doibase 10.1007/JHEP08(2020)111} {\bibfield  {journal} {\bibinfo  {journal}
  {JHEP}\ }\textbf {\bibinfo {volume} {08}},\ \bibinfo {pages} {111} (\bibinfo
  {year} {2020})},\ \Eprint {http://arxiv.org/abs/1909.07318} {arXiv:1909.07318
  [hep-th]} \BibitemShut {NoStop}%
\bibitem [{\citenamefont {Reichert}\ and\ \citenamefont
  {Smirnov}(2020)}]{Reichert:2019car}%
  \BibitemOpen
  \bibfield  {author} {\bibinfo {author} {\bibfnamefont {M.}~\bibnamefont
  {Reichert}}\ and\ \bibinfo {author} {\bibfnamefont {J.}~\bibnamefont
  {Smirnov}},\ }\href {\doibase 10.1103/PhysRevD.101.063015} {\bibfield
  {journal} {\bibinfo  {journal} {Phys. Rev. D}\ }\textbf {\bibinfo {volume}
  {101}},\ \bibinfo {pages} {063015} (\bibinfo {year} {2020})},\ \Eprint
  {http://arxiv.org/abs/1911.00012} {arXiv:1911.00012 [hep-ph]} \BibitemShut
  {NoStop}%
\bibitem [{\citenamefont {Eichhorn}\ and\ \citenamefont
  {Pauly}(2021{\natexlab{a}})}]{Eichhorn:2020kca}%
  \BibitemOpen
  \bibfield  {author} {\bibinfo {author} {\bibfnamefont {A.}~\bibnamefont
  {Eichhorn}}\ and\ \bibinfo {author} {\bibfnamefont {M.}~\bibnamefont
  {Pauly}},\ }\href {\doibase 10.1016/j.physletb.2021.136455} {\bibfield
  {journal} {\bibinfo  {journal} {Phys. Lett. B}\ }\textbf {\bibinfo {volume}
  {819}},\ \bibinfo {pages} {136455} (\bibinfo {year} {2021}{\natexlab{a}})},\
  \Eprint {http://arxiv.org/abs/2005.03661} {arXiv:2005.03661 [hep-ph]}
  \BibitemShut {NoStop}%
\bibitem [{\citenamefont {Eichhorn}\ and\ \citenamefont
  {Pauly}(2021{\natexlab{b}})}]{Eichhorn:2020sbo}%
  \BibitemOpen
  \bibfield  {author} {\bibinfo {author} {\bibfnamefont {A.}~\bibnamefont
  {Eichhorn}}\ and\ \bibinfo {author} {\bibfnamefont {M.}~\bibnamefont
  {Pauly}},\ }\href {\doibase 10.1103/PhysRevD.103.026006} {\bibfield
  {journal} {\bibinfo  {journal} {Phys. Rev. D}\ }\textbf {\bibinfo {volume}
  {103}},\ \bibinfo {pages} {026006} (\bibinfo {year} {2021}{\natexlab{b}})},\
  \Eprint {http://arxiv.org/abs/2009.13543} {arXiv:2009.13543 [hep-th]}
  \BibitemShut {NoStop}%
\bibitem [{\citenamefont {Hamada}\ \emph {et~al.}(2020)\citenamefont {Hamada},
  \citenamefont {Tsumura},\ and\ \citenamefont {Yamada}}]{Hamada:2020vnf}%
  \BibitemOpen
  \bibfield  {author} {\bibinfo {author} {\bibfnamefont {Y.}~\bibnamefont
  {Hamada}}, \bibinfo {author} {\bibfnamefont {K.}~\bibnamefont {Tsumura}}, \
  and\ \bibinfo {author} {\bibfnamefont {M.}~\bibnamefont {Yamada}},\ }\href
  {\doibase 10.1140/epjc/s10052-020-7929-3} {\bibfield  {journal} {\bibinfo
  {journal} {Eur. Phys. J. C}\ }\textbf {\bibinfo {volume} {80}},\ \bibinfo
  {pages} {368} (\bibinfo {year} {2020})},\ \Eprint
  {http://arxiv.org/abs/2002.03666} {arXiv:2002.03666 [hep-ph]} \BibitemShut
  {NoStop}%
\bibitem [{\citenamefont {Kowalska}\ \emph {et~al.}(2021)\citenamefont
  {Kowalska}, \citenamefont {Sessolo},\ and\ \citenamefont
  {Yamamoto}}]{Kowalska:2020gie}%
  \BibitemOpen
  \bibfield  {author} {\bibinfo {author} {\bibfnamefont {K.}~\bibnamefont
  {Kowalska}}, \bibinfo {author} {\bibfnamefont {E.~M.}\ \bibnamefont
  {Sessolo}}, \ and\ \bibinfo {author} {\bibfnamefont {Y.}~\bibnamefont
  {Yamamoto}},\ }\href {\doibase 10.1140/epjc/s10052-021-09072-1} {\bibfield
  {journal} {\bibinfo  {journal} {Eur. Phys. J. C}\ }\textbf {\bibinfo {volume}
  {81}},\ \bibinfo {pages} {272} (\bibinfo {year} {2021})},\ \Eprint
  {http://arxiv.org/abs/2007.03567} {arXiv:2007.03567 [hep-ph]} \BibitemShut
  {NoStop}%
\bibitem [{\citenamefont {Kowalska}\ and\ \citenamefont
  {Sessolo}(2021)}]{Kowalska:2020zve}%
  \BibitemOpen
  \bibfield  {author} {\bibinfo {author} {\bibfnamefont {K.}~\bibnamefont
  {Kowalska}}\ and\ \bibinfo {author} {\bibfnamefont {E.~M.}\ \bibnamefont
  {Sessolo}},\ }\href {\doibase 10.1103/PhysRevD.103.115032} {\bibfield
  {journal} {\bibinfo  {journal} {Phys. Rev. D}\ }\textbf {\bibinfo {volume}
  {103}},\ \bibinfo {pages} {115032} (\bibinfo {year} {2021})},\ \Eprint
  {http://arxiv.org/abs/2012.15200} {arXiv:2012.15200 [hep-ph]} \BibitemShut
  {NoStop}%
\bibitem [{\citenamefont {Held}\ \emph {et~al.}(2022)\citenamefont {Held},
  \citenamefont {Kwapisz},\ and\ \citenamefont {Sartore}}]{Held:2022hnw}%
  \BibitemOpen
  \bibfield  {author} {\bibinfo {author} {\bibfnamefont {A.}~\bibnamefont
  {Held}}, \bibinfo {author} {\bibfnamefont {J.}~\bibnamefont {Kwapisz}}, \
  and\ \bibinfo {author} {\bibfnamefont {L.}~\bibnamefont {Sartore}},\
  }\href@noop {} {\  (\bibinfo {year} {2022})},\ \Eprint
  {http://arxiv.org/abs/2204.03001} {arXiv:2204.03001 [hep-ph]} \BibitemShut
  {NoStop}%
\bibitem [{\citenamefont {Wetterich}\ and\ \citenamefont
  {Yamada}(2017)}]{Wetterich:2016uxm}%
  \BibitemOpen
  \bibfield  {author} {\bibinfo {author} {\bibfnamefont {C.}~\bibnamefont
  {Wetterich}}\ and\ \bibinfo {author} {\bibfnamefont {M.}~\bibnamefont
  {Yamada}},\ }\href {\doibase 10.1016/j.physletb.2017.04.049} {\bibfield
  {journal} {\bibinfo  {journal} {Phys. Lett. B}\ }\textbf {\bibinfo {volume}
  {770}},\ \bibinfo {pages} {268} (\bibinfo {year} {2017})},\ \Eprint
  {http://arxiv.org/abs/1612.03069} {arXiv:1612.03069 [hep-th]} \BibitemShut
  {NoStop}%
\bibitem [{\citenamefont {Weinberg}(1980)}]{Weinberg:1980gg}%
  \BibitemOpen
  \bibfield  {author} {\bibinfo {author} {\bibfnamefont {S.}~\bibnamefont
  {Weinberg}},\ }\enquote {\bibinfo {title} {{ULTRAVIOLET DIVERGENCES IN
  QUANTUM THEORIES OF GRAVITATION}},}\ in\ \href@noop {} {\emph {\bibinfo
  {booktitle} {{General Relativity}: {An Einstein Centenary Survey}}}}\
  (\bibinfo {year} {1980})\ pp.\ \bibinfo {pages} {790--831}\BibitemShut
  {NoStop}%
\bibitem [{\citenamefont {Reuter}(1998)}]{Reuter:1996cp}%
  \BibitemOpen
  \bibfield  {author} {\bibinfo {author} {\bibfnamefont {M.}~\bibnamefont
  {Reuter}},\ }\href {\doibase 10.1103/PhysRevD.57.971} {\bibfield  {journal}
  {\bibinfo  {journal} {Phys. Rev. D}\ }\textbf {\bibinfo {volume} {57}},\
  \bibinfo {pages} {971} (\bibinfo {year} {1998})},\ \Eprint
  {http://arxiv.org/abs/hep-th/9605030} {arXiv:hep-th/9605030} \BibitemShut
  {NoStop}%
\bibitem [{\citenamefont {Reichert}(2020)}]{Reichert:2020mja}%
  \BibitemOpen
  \bibfield  {author} {\bibinfo {author} {\bibfnamefont {M.}~\bibnamefont
  {Reichert}},\ }\href {\doibase 10.22323/1.384.0005} {\bibfield  {journal}
  {\bibinfo  {journal} {PoS}\ }\textbf {\bibinfo {volume} {384}},\ \bibinfo
  {pages} {005} (\bibinfo {year} {2020})}\BibitemShut {NoStop}%
\bibitem [{\citenamefont {Pawlowski}\ and\ \citenamefont
  {Reichert}(2021)}]{Pawlowski:2020qer}%
  \BibitemOpen
  \bibfield  {author} {\bibinfo {author} {\bibfnamefont {J.~M.}\ \bibnamefont
  {Pawlowski}}\ and\ \bibinfo {author} {\bibfnamefont {M.}~\bibnamefont
  {Reichert}},\ }\href {\doibase 10.3389/fphy.2020.551848} {\bibfield
  {journal} {\bibinfo  {journal} {Front. in Phys.}\ }\textbf {\bibinfo {volume}
  {8}},\ \bibinfo {pages} {551848} (\bibinfo {year} {2021})},\ \Eprint
  {http://arxiv.org/abs/2007.10353} {arXiv:2007.10353 [hep-th]} \BibitemShut
  {NoStop}%
\bibitem [{\citenamefont {Percacci}(2017)}]{Percacci:2017fkn}%
  \BibitemOpen
  \bibfield  {author} {\bibinfo {author} {\bibfnamefont {R.}~\bibnamefont
  {Percacci}},\ }\href {\doibase 10.1142/10369} {\emph {\bibinfo {title} {{An
  Introduction to Covariant Quantum Gravity and Asymptotic Safety}}}},\
  \bibinfo {series} {100 Years of General Relativity}, Vol.~\bibinfo {volume}
  {3}\ (\bibinfo  {publisher} {World Scientific},\ \bibinfo {year}
  {2017})\BibitemShut {NoStop}%
\bibitem [{\citenamefont {Reuter}\ and\ \citenamefont
  {Saueressig}(2019)}]{Reuter:2019byg}%
  \BibitemOpen
  \bibfield  {author} {\bibinfo {author} {\bibfnamefont {M.}~\bibnamefont
  {Reuter}}\ and\ \bibinfo {author} {\bibfnamefont {F.}~\bibnamefont
  {Saueressig}},\ }\href@noop {} {\emph {\bibinfo {title} {{Quantum Gravity and
  the Functional Renormalization Group}: {The Road towards Asymptotic
  Safety}}}}\ (\bibinfo  {publisher} {Cambridge University Press},\ \bibinfo
  {year} {2019})\BibitemShut {NoStop}%
\bibitem [{\citenamefont {Bonanno}\ \emph {et~al.}(2020)\citenamefont
  {Bonanno}, \citenamefont {Eichhorn}, \citenamefont {Gies}, \citenamefont
  {Pawlowski}, \citenamefont {Percacci}, \citenamefont {Reuter}, \citenamefont
  {Saueressig},\ and\ \citenamefont {Vacca}}]{Bonanno:2020bil}%
  \BibitemOpen
  \bibfield  {author} {\bibinfo {author} {\bibfnamefont {A.}~\bibnamefont
  {Bonanno}}, \bibinfo {author} {\bibfnamefont {A.}~\bibnamefont {Eichhorn}},
  \bibinfo {author} {\bibfnamefont {H.}~\bibnamefont {Gies}}, \bibinfo {author}
  {\bibfnamefont {J.~M.}\ \bibnamefont {Pawlowski}}, \bibinfo {author}
  {\bibfnamefont {R.}~\bibnamefont {Percacci}}, \bibinfo {author}
  {\bibfnamefont {M.}~\bibnamefont {Reuter}}, \bibinfo {author} {\bibfnamefont
  {F.}~\bibnamefont {Saueressig}}, \ and\ \bibinfo {author} {\bibfnamefont
  {G.~P.}\ \bibnamefont {Vacca}},\ }\href {\doibase 10.3389/fphy.2020.00269}
  {\bibfield  {journal} {\bibinfo  {journal} {Front. in Phys.}\ }\textbf
  {\bibinfo {volume} {8}},\ \bibinfo {pages} {269} (\bibinfo {year} {2020})},\
  \Eprint {http://arxiv.org/abs/2004.06810} {arXiv:2004.06810 [gr-qc]}
  \BibitemShut {NoStop}%
\bibitem [{\citenamefont {Draper}\ \emph {et~al.}(2020)\citenamefont {Draper},
  \citenamefont {Knorr}, \citenamefont {Ripken},\ and\ \citenamefont
  {Saueressig}}]{Draper:2020bop}%
  \BibitemOpen
  \bibfield  {author} {\bibinfo {author} {\bibfnamefont {T.}~\bibnamefont
  {Draper}}, \bibinfo {author} {\bibfnamefont {B.}~\bibnamefont {Knorr}},
  \bibinfo {author} {\bibfnamefont {C.}~\bibnamefont {Ripken}}, \ and\ \bibinfo
  {author} {\bibfnamefont {F.}~\bibnamefont {Saueressig}},\ }\href {\doibase
  10.1103/PhysRevLett.125.181301} {\bibfield  {journal} {\bibinfo  {journal}
  {Phys. Rev. Lett.}\ }\textbf {\bibinfo {volume} {125}},\ \bibinfo {pages}
  {181301} (\bibinfo {year} {2020})},\ \Eprint
  {http://arxiv.org/abs/2007.00733} {arXiv:2007.00733 [hep-th]} \BibitemShut
  {NoStop}%
\bibitem [{\citenamefont {Platania}\ and\ \citenamefont
  {Wetterich}(2020)}]{Platania:2020knd}%
  \BibitemOpen
  \bibfield  {author} {\bibinfo {author} {\bibfnamefont {A.}~\bibnamefont
  {Platania}}\ and\ \bibinfo {author} {\bibfnamefont {C.}~\bibnamefont
  {Wetterich}},\ }\href {\doibase 10.1016/j.physletb.2020.135911} {\bibfield
  {journal} {\bibinfo  {journal} {Phys. Lett. B}\ }\textbf {\bibinfo {volume}
  {811}},\ \bibinfo {pages} {135911} (\bibinfo {year} {2020})},\ \Eprint
  {http://arxiv.org/abs/2009.06637} {arXiv:2009.06637 [hep-th]} \BibitemShut
  {NoStop}%
\bibitem [{\citenamefont {Fehre}\ \emph {et~al.}(2021)\citenamefont {Fehre},
  \citenamefont {Litim}, \citenamefont {Pawlowski},\ and\ \citenamefont
  {Reichert}}]{Fehre:2021eob}%
  \BibitemOpen
  \bibfield  {author} {\bibinfo {author} {\bibfnamefont {J.}~\bibnamefont
  {Fehre}}, \bibinfo {author} {\bibfnamefont {D.~F.}\ \bibnamefont {Litim}},
  \bibinfo {author} {\bibfnamefont {J.~M.}\ \bibnamefont {Pawlowski}}, \ and\
  \bibinfo {author} {\bibfnamefont {M.}~\bibnamefont {Reichert}},\ }\href@noop
  {} {\  (\bibinfo {year} {2021})},\ \Eprint {http://arxiv.org/abs/2111.13232}
  {arXiv:2111.13232 [hep-th]} \BibitemShut {NoStop}%
\bibitem [{\citenamefont {Don\`a}\ \emph {et~al.}(2014)\citenamefont {Don\`a},
  \citenamefont {Eichhorn},\ and\ \citenamefont {Percacci}}]{Dona:2013qba}%
  \BibitemOpen
  \bibfield  {author} {\bibinfo {author} {\bibfnamefont {P.}~\bibnamefont
  {Don\`a}}, \bibinfo {author} {\bibfnamefont {A.}~\bibnamefont {Eichhorn}}, \
  and\ \bibinfo {author} {\bibfnamefont {R.}~\bibnamefont {Percacci}},\ }\href
  {\doibase 10.1103/PhysRevD.89.084035} {\bibfield  {journal} {\bibinfo
  {journal} {Phys. Rev. D}\ }\textbf {\bibinfo {volume} {89}},\ \bibinfo
  {pages} {084035} (\bibinfo {year} {2014})},\ \Eprint
  {http://arxiv.org/abs/1311.2898} {arXiv:1311.2898 [hep-th]} \BibitemShut
  {NoStop}%
\bibitem [{\citenamefont {Meibohm}\ \emph {et~al.}(2016)\citenamefont
  {Meibohm}, \citenamefont {Pawlowski},\ and\ \citenamefont
  {Reichert}}]{Meibohm:2015twa}%
  \BibitemOpen
  \bibfield  {author} {\bibinfo {author} {\bibfnamefont {J.}~\bibnamefont
  {Meibohm}}, \bibinfo {author} {\bibfnamefont {J.~M.}\ \bibnamefont
  {Pawlowski}}, \ and\ \bibinfo {author} {\bibfnamefont {M.}~\bibnamefont
  {Reichert}},\ }\href {\doibase 10.1103/PhysRevD.93.084035} {\bibfield
  {journal} {\bibinfo  {journal} {Phys. Rev. D}\ }\textbf {\bibinfo {volume}
  {93}},\ \bibinfo {pages} {084035} (\bibinfo {year} {2016})},\ \Eprint
  {http://arxiv.org/abs/1510.07018} {arXiv:1510.07018 [hep-th]} \BibitemShut
  {NoStop}%
\bibitem [{\citenamefont {Biemans}\ \emph {et~al.}(2017)\citenamefont
  {Biemans}, \citenamefont {Platania},\ and\ \citenamefont
  {Saueressig}}]{Biemans:2017zca}%
  \BibitemOpen
  \bibfield  {author} {\bibinfo {author} {\bibfnamefont {J.}~\bibnamefont
  {Biemans}}, \bibinfo {author} {\bibfnamefont {A.}~\bibnamefont {Platania}}, \
  and\ \bibinfo {author} {\bibfnamefont {F.}~\bibnamefont {Saueressig}},\
  }\href {\doibase 10.1007/JHEP05(2017)093} {\bibfield  {journal} {\bibinfo
  {journal} {JHEP}\ }\textbf {\bibinfo {volume} {05}},\ \bibinfo {pages} {093}
  (\bibinfo {year} {2017})},\ \Eprint {http://arxiv.org/abs/1702.06539}
  {arXiv:1702.06539 [hep-th]} \BibitemShut {NoStop}%
\bibitem [{\citenamefont {Alkofer}\ and\ \citenamefont
  {Saueressig}(2018)}]{Alkofer:2018fxj}%
  \BibitemOpen
  \bibfield  {author} {\bibinfo {author} {\bibfnamefont {N.}~\bibnamefont
  {Alkofer}}\ and\ \bibinfo {author} {\bibfnamefont {F.}~\bibnamefont
  {Saueressig}},\ }\href {\doibase 10.1016/j.aop.2018.07.017} {\bibfield
  {journal} {\bibinfo  {journal} {Annals Phys.}\ }\textbf {\bibinfo {volume}
  {396}},\ \bibinfo {pages} {173} (\bibinfo {year} {2018})},\ \Eprint
  {http://arxiv.org/abs/1802.00498} {arXiv:1802.00498 [hep-th]} \BibitemShut
  {NoStop}%
\bibitem [{\citenamefont {Wetterich}\ and\ \citenamefont
  {Yamada}(2019)}]{Wetterich:2019zdo}%
  \BibitemOpen
  \bibfield  {author} {\bibinfo {author} {\bibfnamefont {C.}~\bibnamefont
  {Wetterich}}\ and\ \bibinfo {author} {\bibfnamefont {M.}~\bibnamefont
  {Yamada}},\ }\href {\doibase 10.1103/PhysRevD.100.066017} {\bibfield
  {journal} {\bibinfo  {journal} {Phys. Rev. D}\ }\textbf {\bibinfo {volume}
  {100}},\ \bibinfo {pages} {066017} (\bibinfo {year} {2019})},\ \Eprint
  {http://arxiv.org/abs/1906.01721} {arXiv:1906.01721 [hep-th]} \BibitemShut
  {NoStop}%
\bibitem [{\citenamefont {Sen}\ \emph {et~al.}(2022)\citenamefont {Sen},
  \citenamefont {Wetterich},\ and\ \citenamefont {Yamada}}]{Sen:2021ffc}%
  \BibitemOpen
  \bibfield  {author} {\bibinfo {author} {\bibfnamefont {S.}~\bibnamefont
  {Sen}}, \bibinfo {author} {\bibfnamefont {C.}~\bibnamefont {Wetterich}}, \
  and\ \bibinfo {author} {\bibfnamefont {M.}~\bibnamefont {Yamada}},\ }\href
  {\doibase 10.1007/JHEP03(2022)130} {\bibfield  {journal} {\bibinfo  {journal}
  {JHEP}\ }\textbf {\bibinfo {volume} {03}},\ \bibinfo {pages} {130} (\bibinfo
  {year} {2022})},\ \Eprint {http://arxiv.org/abs/2111.04696} {arXiv:2111.04696
  [hep-th]} \BibitemShut {NoStop}%
\bibitem [{\citenamefont {Narain}\ and\ \citenamefont
  {Percacci}(2010)}]{Narain:2009fy}%
  \BibitemOpen
  \bibfield  {author} {\bibinfo {author} {\bibfnamefont {G.}~\bibnamefont
  {Narain}}\ and\ \bibinfo {author} {\bibfnamefont {R.}~\bibnamefont
  {Percacci}},\ }\href {\doibase 10.1088/0264-9381/27/7/075001} {\bibfield
  {journal} {\bibinfo  {journal} {Class. Quant. Grav.}\ }\textbf {\bibinfo
  {volume} {27}},\ \bibinfo {pages} {075001} (\bibinfo {year} {2010})},\
  \Eprint {http://arxiv.org/abs/0911.0386} {arXiv:0911.0386 [hep-th]}
  \BibitemShut {NoStop}%
\bibitem [{\citenamefont {Folkerts}\ \emph {et~al.}(2012)\citenamefont
  {Folkerts}, \citenamefont {Litim},\ and\ \citenamefont
  {Pawlowski}}]{Folkerts:2011jz}%
  \BibitemOpen
  \bibfield  {author} {\bibinfo {author} {\bibfnamefont {S.}~\bibnamefont
  {Folkerts}}, \bibinfo {author} {\bibfnamefont {D.~F.}\ \bibnamefont {Litim}},
  \ and\ \bibinfo {author} {\bibfnamefont {J.~M.}\ \bibnamefont {Pawlowski}},\
  }\href {\doibase 10.1016/j.physletb.2012.02.002} {\bibfield  {journal}
  {\bibinfo  {journal} {Phys. Lett. B}\ }\textbf {\bibinfo {volume} {709}},\
  \bibinfo {pages} {234} (\bibinfo {year} {2012})},\ \Eprint
  {http://arxiv.org/abs/1101.5552} {arXiv:1101.5552 [hep-th]} \BibitemShut
  {NoStop}%
\bibitem [{\citenamefont {Eichhorn}\ and\ \citenamefont
  {Gies}(2011)}]{Eichhorn:2011pc}%
  \BibitemOpen
  \bibfield  {author} {\bibinfo {author} {\bibfnamefont {A.}~\bibnamefont
  {Eichhorn}}\ and\ \bibinfo {author} {\bibfnamefont {H.}~\bibnamefont
  {Gies}},\ }\href {\doibase 10.1088/1367-2630/13/12/125012} {\bibfield
  {journal} {\bibinfo  {journal} {New J. Phys.}\ }\textbf {\bibinfo {volume}
  {13}},\ \bibinfo {pages} {125012} (\bibinfo {year} {2011})},\ \Eprint
  {http://arxiv.org/abs/1104.5366} {arXiv:1104.5366 [hep-th]} \BibitemShut
  {NoStop}%
\bibitem [{\citenamefont {Eichhorn}(2012)}]{Eichhorn:2012va}%
  \BibitemOpen
  \bibfield  {author} {\bibinfo {author} {\bibfnamefont {A.}~\bibnamefont
  {Eichhorn}},\ }\href {\doibase 10.1103/PhysRevD.86.105021} {\bibfield
  {journal} {\bibinfo  {journal} {Phys. Rev. D}\ }\textbf {\bibinfo {volume}
  {86}},\ \bibinfo {pages} {105021} (\bibinfo {year} {2012})},\ \Eprint
  {http://arxiv.org/abs/1204.0965} {arXiv:1204.0965 [gr-qc]} \BibitemShut
  {NoStop}%
\bibitem [{\citenamefont {Percacci}\ and\ \citenamefont
  {Vacca}(2015)}]{Percacci:2015wwa}%
  \BibitemOpen
  \bibfield  {author} {\bibinfo {author} {\bibfnamefont {R.}~\bibnamefont
  {Percacci}}\ and\ \bibinfo {author} {\bibfnamefont {G.~P.}\ \bibnamefont
  {Vacca}},\ }\href {\doibase 10.1140/epjc/s10052-015-3410-0} {\bibfield
  {journal} {\bibinfo  {journal} {Eur. Phys. J. C}\ }\textbf {\bibinfo {volume}
  {75}},\ \bibinfo {pages} {188} (\bibinfo {year} {2015})},\ \Eprint
  {http://arxiv.org/abs/1501.00888} {arXiv:1501.00888 [hep-th]} \BibitemShut
  {NoStop}%
\bibitem [{\citenamefont {Eichhorn}\ and\ \citenamefont
  {Lippoldt}(2017)}]{Eichhorn:2016vvy}%
  \BibitemOpen
  \bibfield  {author} {\bibinfo {author} {\bibfnamefont {A.}~\bibnamefont
  {Eichhorn}}\ and\ \bibinfo {author} {\bibfnamefont {S.}~\bibnamefont
  {Lippoldt}},\ }\href {\doibase 10.1016/j.physletb.2017.01.064} {\bibfield
  {journal} {\bibinfo  {journal} {Phys. Lett. B}\ }\textbf {\bibinfo {volume}
  {767}},\ \bibinfo {pages} {142} (\bibinfo {year} {2017})},\ \Eprint
  {http://arxiv.org/abs/1611.05878} {arXiv:1611.05878 [gr-qc]} \BibitemShut
  {NoStop}%
\bibitem [{\citenamefont {Eichhorn}\ \emph
  {et~al.}(2018{\natexlab{c}})\citenamefont {Eichhorn}, \citenamefont
  {Lippoldt},\ and\ \citenamefont {Skrinjar}}]{Eichhorn:2017sok}%
  \BibitemOpen
  \bibfield  {author} {\bibinfo {author} {\bibfnamefont {A.}~\bibnamefont
  {Eichhorn}}, \bibinfo {author} {\bibfnamefont {S.}~\bibnamefont {Lippoldt}},
  \ and\ \bibinfo {author} {\bibfnamefont {V.}~\bibnamefont {Skrinjar}},\
  }\href {\doibase 10.1103/PhysRevD.97.026002} {\bibfield  {journal} {\bibinfo
  {journal} {Phys. Rev. D}\ }\textbf {\bibinfo {volume} {97}},\ \bibinfo
  {pages} {026002} (\bibinfo {year} {2018}{\natexlab{c}})},\ \Eprint
  {http://arxiv.org/abs/1710.03005} {arXiv:1710.03005 [hep-th]} \BibitemShut
  {NoStop}%
\bibitem [{\citenamefont {Christiansen}\ and\ \citenamefont
  {Eichhorn}(2017)}]{Christiansen:2017gtg}%
  \BibitemOpen
  \bibfield  {author} {\bibinfo {author} {\bibfnamefont {N.}~\bibnamefont
  {Christiansen}}\ and\ \bibinfo {author} {\bibfnamefont {A.}~\bibnamefont
  {Eichhorn}},\ }\href {\doibase 10.1016/j.physletb.2017.04.047} {\bibfield
  {journal} {\bibinfo  {journal} {Phys. Lett. B}\ }\textbf {\bibinfo {volume}
  {770}},\ \bibinfo {pages} {154} (\bibinfo {year} {2017})},\ \Eprint
  {http://arxiv.org/abs/1702.07724} {arXiv:1702.07724 [hep-th]} \BibitemShut
  {NoStop}%
\bibitem [{\citenamefont {Christiansen}\ \emph {et~al.}(2018)\citenamefont
  {Christiansen}, \citenamefont {Litim}, \citenamefont {Pawlowski},\ and\
  \citenamefont {Reichert}}]{Christiansen:2017cxa}%
  \BibitemOpen
  \bibfield  {author} {\bibinfo {author} {\bibfnamefont {N.}~\bibnamefont
  {Christiansen}}, \bibinfo {author} {\bibfnamefont {D.~F.}\ \bibnamefont
  {Litim}}, \bibinfo {author} {\bibfnamefont {J.~M.}\ \bibnamefont
  {Pawlowski}}, \ and\ \bibinfo {author} {\bibfnamefont {M.}~\bibnamefont
  {Reichert}},\ }\href {\doibase 10.1103/PhysRevD.97.106012} {\bibfield
  {journal} {\bibinfo  {journal} {Phys. Rev. D}\ }\textbf {\bibinfo {volume}
  {97}},\ \bibinfo {pages} {106012} (\bibinfo {year} {2018})},\ \Eprint
  {http://arxiv.org/abs/1710.04669} {arXiv:1710.04669 [hep-th]} \BibitemShut
  {NoStop}%
\bibitem [{\citenamefont {Eichhorn}\ and\ \citenamefont
  {Held}(2017)}]{Eichhorn:2017eht}%
  \BibitemOpen
  \bibfield  {author} {\bibinfo {author} {\bibfnamefont {A.}~\bibnamefont
  {Eichhorn}}\ and\ \bibinfo {author} {\bibfnamefont {A.}~\bibnamefont
  {Held}},\ }\href {\doibase 10.1103/PhysRevD.96.086025} {\bibfield  {journal}
  {\bibinfo  {journal} {Phys. Rev. D}\ }\textbf {\bibinfo {volume} {96}},\
  \bibinfo {pages} {086025} (\bibinfo {year} {2017})},\ \Eprint
  {http://arxiv.org/abs/1705.02342} {arXiv:1705.02342 [gr-qc]} \BibitemShut
  {NoStop}%
\bibitem [{\citenamefont {Eichhorn}\ \emph
  {et~al.}(2019{\natexlab{a}})\citenamefont {Eichhorn}, \citenamefont
  {Lippoldt},\ and\ \citenamefont {Schiffer}}]{Eichhorn:2018nda}%
  \BibitemOpen
  \bibfield  {author} {\bibinfo {author} {\bibfnamefont {A.}~\bibnamefont
  {Eichhorn}}, \bibinfo {author} {\bibfnamefont {S.}~\bibnamefont {Lippoldt}},
  \ and\ \bibinfo {author} {\bibfnamefont {M.}~\bibnamefont {Schiffer}},\
  }\href {\doibase 10.1103/PhysRevD.99.086002} {\bibfield  {journal} {\bibinfo
  {journal} {Phys. Rev. D}\ }\textbf {\bibinfo {volume} {99}},\ \bibinfo
  {pages} {086002} (\bibinfo {year} {2019}{\natexlab{a}})},\ \Eprint
  {http://arxiv.org/abs/1812.08782} {arXiv:1812.08782 [hep-th]} \BibitemShut
  {NoStop}%
\bibitem [{\citenamefont {Pawlowski}\ \emph {et~al.}(2019)\citenamefont
  {Pawlowski}, \citenamefont {Reichert}, \citenamefont {Wetterich},\ and\
  \citenamefont {Yamada}}]{Pawlowski:2018ixd}%
  \BibitemOpen
  \bibfield  {author} {\bibinfo {author} {\bibfnamefont {J.~M.}\ \bibnamefont
  {Pawlowski}}, \bibinfo {author} {\bibfnamefont {M.}~\bibnamefont {Reichert}},
  \bibinfo {author} {\bibfnamefont {C.}~\bibnamefont {Wetterich}}, \ and\
  \bibinfo {author} {\bibfnamefont {M.}~\bibnamefont {Yamada}},\ }\href
  {\doibase 10.1103/PhysRevD.99.086010} {\bibfield  {journal} {\bibinfo
  {journal} {Phys. Rev. D}\ }\textbf {\bibinfo {volume} {99}},\ \bibinfo
  {pages} {086010} (\bibinfo {year} {2019})},\ \Eprint
  {http://arxiv.org/abs/1811.11706} {arXiv:1811.11706 [hep-th]} \BibitemShut
  {NoStop}%
\bibitem [{\citenamefont {Eichhorn}\ \emph
  {et~al.}(2019{\natexlab{b}})\citenamefont {Eichhorn}, \citenamefont
  {Lippoldt}, \citenamefont {Pawlowski}, \citenamefont {Reichert},\ and\
  \citenamefont {Schiffer}}]{Eichhorn:2018ydy}%
  \BibitemOpen
  \bibfield  {author} {\bibinfo {author} {\bibfnamefont {A.}~\bibnamefont
  {Eichhorn}}, \bibinfo {author} {\bibfnamefont {S.}~\bibnamefont {Lippoldt}},
  \bibinfo {author} {\bibfnamefont {J.~M.}\ \bibnamefont {Pawlowski}}, \bibinfo
  {author} {\bibfnamefont {M.}~\bibnamefont {Reichert}}, \ and\ \bibinfo
  {author} {\bibfnamefont {M.}~\bibnamefont {Schiffer}},\ }\href {\doibase
  10.1016/j.physletb.2019.01.071} {\bibfield  {journal} {\bibinfo  {journal}
  {Phys. Lett. B}\ }\textbf {\bibinfo {volume} {792}},\ \bibinfo {pages} {310}
  (\bibinfo {year} {2019}{\natexlab{b}})},\ \Eprint
  {http://arxiv.org/abs/1810.02828} {arXiv:1810.02828 [hep-th]} \BibitemShut
  {NoStop}%
\bibitem [{\citenamefont {Eichhorn}\ \emph
  {et~al.}(2018{\natexlab{d}})\citenamefont {Eichhorn}, \citenamefont {Labus},
  \citenamefont {Pawlowski},\ and\ \citenamefont
  {Reichert}}]{Eichhorn:2018akn}%
  \BibitemOpen
  \bibfield  {author} {\bibinfo {author} {\bibfnamefont {A.}~\bibnamefont
  {Eichhorn}}, \bibinfo {author} {\bibfnamefont {P.}~\bibnamefont {Labus}},
  \bibinfo {author} {\bibfnamefont {J.~M.}\ \bibnamefont {Pawlowski}}, \ and\
  \bibinfo {author} {\bibfnamefont {M.}~\bibnamefont {Reichert}},\ }\href
  {\doibase 10.21468/SciPostPhys.5.4.031} {\bibfield  {journal} {\bibinfo
  {journal} {SciPost Phys.}\ }\textbf {\bibinfo {volume} {5}},\ \bibinfo
  {pages} {031} (\bibinfo {year} {2018}{\natexlab{d}})},\ \Eprint
  {http://arxiv.org/abs/1804.00012} {arXiv:1804.00012 [hep-th]} \BibitemShut
  {NoStop}%
\bibitem [{\citenamefont {Gies}\ and\ \citenamefont
  {Martini}(2018)}]{Gies:2018jnv}%
  \BibitemOpen
  \bibfield  {author} {\bibinfo {author} {\bibfnamefont {H.}~\bibnamefont
  {Gies}}\ and\ \bibinfo {author} {\bibfnamefont {R.}~\bibnamefont {Martini}},\
  }\href {\doibase 10.1103/PhysRevD.97.085017} {\bibfield  {journal} {\bibinfo
  {journal} {Phys. Rev. D}\ }\textbf {\bibinfo {volume} {97}},\ \bibinfo
  {pages} {085017} (\bibinfo {year} {2018})},\ \Eprint
  {http://arxiv.org/abs/1802.02865} {arXiv:1802.02865 [hep-th]} \BibitemShut
  {NoStop}%
\bibitem [{\citenamefont {Eichhorn}\ and\ \citenamefont
  {Schiffer}(2019)}]{Eichhorn:2019yzm}%
  \BibitemOpen
  \bibfield  {author} {\bibinfo {author} {\bibfnamefont {A.}~\bibnamefont
  {Eichhorn}}\ and\ \bibinfo {author} {\bibfnamefont {M.}~\bibnamefont
  {Schiffer}},\ }\href {\doibase 10.1016/j.physletb.2019.05.005} {\bibfield
  {journal} {\bibinfo  {journal} {Phys. Lett. B}\ }\textbf {\bibinfo {volume}
  {793}},\ \bibinfo {pages} {383} (\bibinfo {year} {2019})},\ \Eprint
  {http://arxiv.org/abs/1902.06479} {arXiv:1902.06479 [hep-th]} \BibitemShut
  {NoStop}%
\bibitem [{\citenamefont {De~Brito}\ \emph
  {et~al.}(2019{\natexlab{a}})\citenamefont {De~Brito}, \citenamefont {Hamada},
  \citenamefont {Pereira},\ and\ \citenamefont {Yamada}}]{DeBrito:2019rrh}%
  \BibitemOpen
  \bibfield  {author} {\bibinfo {author} {\bibfnamefont {G.~P.}\ \bibnamefont
  {De~Brito}}, \bibinfo {author} {\bibfnamefont {Y.}~\bibnamefont {Hamada}},
  \bibinfo {author} {\bibfnamefont {A.~D.}\ \bibnamefont {Pereira}}, \ and\
  \bibinfo {author} {\bibfnamefont {M.}~\bibnamefont {Yamada}},\ }\href
  {\doibase 10.1007/JHEP08(2019)142} {\bibfield  {journal} {\bibinfo  {journal}
  {JHEP}\ }\textbf {\bibinfo {volume} {08}},\ \bibinfo {pages} {142} (\bibinfo
  {year} {2019}{\natexlab{a}})},\ \Eprint {http://arxiv.org/abs/1905.11114}
  {arXiv:1905.11114 [hep-th]} \BibitemShut {NoStop}%
\bibitem [{\citenamefont {Wetterich}(2021)}]{Wetterich:2019rsn}%
  \BibitemOpen
  \bibfield  {author} {\bibinfo {author} {\bibfnamefont {C.}~\bibnamefont
  {Wetterich}},\ }\href {\doibase 10.3390/universe7020045} {\bibfield
  {journal} {\bibinfo  {journal} {Universe}\ }\textbf {\bibinfo {volume} {7}},\
  \bibinfo {pages} {45} (\bibinfo {year} {2021})},\ \Eprint
  {http://arxiv.org/abs/1911.06100} {arXiv:1911.06100 [hep-th]} \BibitemShut
  {NoStop}%
\bibitem [{\citenamefont {Hamada}\ \emph {et~al.}(2021)\citenamefont {Hamada},
  \citenamefont {Pawlowski},\ and\ \citenamefont {Yamada}}]{Hamada:2020mug}%
  \BibitemOpen
  \bibfield  {author} {\bibinfo {author} {\bibfnamefont {Y.}~\bibnamefont
  {Hamada}}, \bibinfo {author} {\bibfnamefont {J.~M.}\ \bibnamefont
  {Pawlowski}}, \ and\ \bibinfo {author} {\bibfnamefont {M.}~\bibnamefont
  {Yamada}},\ }\href {\doibase 10.1103/PhysRevD.103.106016} {\bibfield
  {journal} {\bibinfo  {journal} {Phys. Rev. D}\ }\textbf {\bibinfo {volume}
  {103}},\ \bibinfo {pages} {106016} (\bibinfo {year} {2021})},\ \Eprint
  {http://arxiv.org/abs/2009.08728} {arXiv:2009.08728 [hep-th]} \BibitemShut
  {NoStop}%
\bibitem [{\citenamefont {de~Brito}\ \emph
  {et~al.}(2021{\natexlab{a}})\citenamefont {de~Brito}, \citenamefont
  {Eichhorn},\ and\ \citenamefont {Schiffer}}]{deBrito:2020dta}%
  \BibitemOpen
  \bibfield  {author} {\bibinfo {author} {\bibfnamefont {G.~P.}\ \bibnamefont
  {de~Brito}}, \bibinfo {author} {\bibfnamefont {A.}~\bibnamefont {Eichhorn}},
  \ and\ \bibinfo {author} {\bibfnamefont {M.}~\bibnamefont {Schiffer}},\
  }\href {\doibase 10.1016/j.physletb.2021.136128} {\bibfield  {journal}
  {\bibinfo  {journal} {Phys. Lett. B}\ }\textbf {\bibinfo {volume} {815}},\
  \bibinfo {pages} {136128} (\bibinfo {year} {2021}{\natexlab{a}})},\ \Eprint
  {http://arxiv.org/abs/2010.00605} {arXiv:2010.00605 [hep-th]} \BibitemShut
  {NoStop}%
\bibitem [{\citenamefont {Daas}\ \emph {et~al.}(2021)\citenamefont {Daas},
  \citenamefont {Oosters}, \citenamefont {Saueressig},\ and\ \citenamefont
  {Wang}}]{Daas:2021abx}%
  \BibitemOpen
  \bibfield  {author} {\bibinfo {author} {\bibfnamefont {J.}~\bibnamefont
  {Daas}}, \bibinfo {author} {\bibfnamefont {W.}~\bibnamefont {Oosters}},
  \bibinfo {author} {\bibfnamefont {F.}~\bibnamefont {Saueressig}}, \ and\
  \bibinfo {author} {\bibfnamefont {J.}~\bibnamefont {Wang}},\ }\href {\doibase
  10.3390/universe7080306} {\bibfield  {journal} {\bibinfo  {journal}
  {Universe}\ }\textbf {\bibinfo {volume} {7}},\ \bibinfo {pages} {306}
  (\bibinfo {year} {2021})},\ \Eprint {http://arxiv.org/abs/2107.01071}
  {arXiv:2107.01071 [hep-th]} \BibitemShut {NoStop}%
\bibitem [{\citenamefont {Gies}\ and\ \citenamefont
  {Salek}(2021)}]{Gies:2021upb}%
  \BibitemOpen
  \bibfield  {author} {\bibinfo {author} {\bibfnamefont {H.}~\bibnamefont
  {Gies}}\ and\ \bibinfo {author} {\bibfnamefont {A.~S.}\ \bibnamefont
  {Salek}},\ }\href {\doibase 10.1103/PhysRevD.103.125027} {\bibfield
  {journal} {\bibinfo  {journal} {Phys. Rev. D}\ }\textbf {\bibinfo {volume}
  {103}},\ \bibinfo {pages} {125027} (\bibinfo {year} {2021})},\ \Eprint
  {http://arxiv.org/abs/2103.05542} {arXiv:2103.05542 [hep-th]} \BibitemShut
  {NoStop}%
\bibitem [{\citenamefont {de~Brito}\ \emph
  {et~al.}(2021{\natexlab{b}})\citenamefont {de~Brito}, \citenamefont
  {Eichhorn},\ and\ \citenamefont {Santos}}]{deBrito:2021pyi}%
  \BibitemOpen
  \bibfield  {author} {\bibinfo {author} {\bibfnamefont {G.~P.}\ \bibnamefont
  {de~Brito}}, \bibinfo {author} {\bibfnamefont {A.}~\bibnamefont {Eichhorn}},
  \ and\ \bibinfo {author} {\bibfnamefont {R.~R. L.~d.}\ \bibnamefont
  {Santos}},\ }\href {\doibase 10.1007/JHEP11(2021)110} {\bibfield  {journal}
  {\bibinfo  {journal} {JHEP}\ }\textbf {\bibinfo {volume} {11}},\ \bibinfo
  {pages} {110} (\bibinfo {year} {2021}{\natexlab{b}})},\ \Eprint
  {http://arxiv.org/abs/2107.03839} {arXiv:2107.03839 [gr-qc]} \BibitemShut
  {NoStop}%
\bibitem [{\citenamefont {Laporte}\ \emph {et~al.}(2021)\citenamefont
  {Laporte}, \citenamefont {Pereira}, \citenamefont {Saueressig},\ and\
  \citenamefont {Wang}}]{Laporte:2021kyp}%
  \BibitemOpen
  \bibfield  {author} {\bibinfo {author} {\bibfnamefont {C.}~\bibnamefont
  {Laporte}}, \bibinfo {author} {\bibfnamefont {A.~D.}\ \bibnamefont
  {Pereira}}, \bibinfo {author} {\bibfnamefont {F.}~\bibnamefont {Saueressig}},
  \ and\ \bibinfo {author} {\bibfnamefont {J.}~\bibnamefont {Wang}},\ }\href
  {\doibase 10.1007/JHEP12(2021)001} {\bibfield  {journal} {\bibinfo  {journal}
  {JHEP}\ }\textbf {\bibinfo {volume} {12}},\ \bibinfo {pages} {001} (\bibinfo
  {year} {2021})},\ \Eprint {http://arxiv.org/abs/2110.09566} {arXiv:2110.09566
  [hep-th]} \BibitemShut {NoStop}%
\bibitem [{\citenamefont {Ohta}\ and\ \citenamefont
  {Yamada}(2022)}]{Ohta:2021bkc}%
  \BibitemOpen
  \bibfield  {author} {\bibinfo {author} {\bibfnamefont {N.}~\bibnamefont
  {Ohta}}\ and\ \bibinfo {author} {\bibfnamefont {M.}~\bibnamefont {Yamada}},\
  }\href {\doibase 10.1103/PhysRevD.105.026013} {\bibfield  {journal} {\bibinfo
   {journal} {Phys. Rev. D}\ }\textbf {\bibinfo {volume} {105}},\ \bibinfo
  {pages} {026013} (\bibinfo {year} {2022})},\ \Eprint
  {http://arxiv.org/abs/2110.08594} {arXiv:2110.08594 [hep-th]} \BibitemShut
  {NoStop}%
\bibitem [{\citenamefont {Eichhorn}\ \emph
  {et~al.}(2021{\natexlab{a}})\citenamefont {Eichhorn}, \citenamefont
  {Kwapisz},\ and\ \citenamefont {Schiffer}}]{Eichhorn:2021qet}%
  \BibitemOpen
  \bibfield  {author} {\bibinfo {author} {\bibfnamefont {A.}~\bibnamefont
  {Eichhorn}}, \bibinfo {author} {\bibfnamefont {J.~H.}\ \bibnamefont
  {Kwapisz}}, \ and\ \bibinfo {author} {\bibfnamefont {M.}~\bibnamefont
  {Schiffer}},\ }\href@noop {} {\  (\bibinfo {year} {2021}{\natexlab{a}})},\
  \Eprint {http://arxiv.org/abs/2112.09772} {arXiv:2112.09772 [gr-qc]}
  \BibitemShut {NoStop}%
\bibitem [{\citenamefont {Donoghue}(2020)}]{Donoghue:2019clr}%
  \BibitemOpen
  \bibfield  {author} {\bibinfo {author} {\bibfnamefont {J.~F.}\ \bibnamefont
  {Donoghue}},\ }\href {\doibase 10.3389/fphy.2020.00056} {\bibfield  {journal}
  {\bibinfo  {journal} {Front. in Phys.}\ }\textbf {\bibinfo {volume} {8}},\
  \bibinfo {pages} {56} (\bibinfo {year} {2020})},\ \Eprint
  {http://arxiv.org/abs/1911.02967} {arXiv:1911.02967 [hep-th]} \BibitemShut
  {NoStop}%
\bibitem [{\citenamefont {Daum}\ \emph {et~al.}(2010)\citenamefont {Daum},
  \citenamefont {Harst},\ and\ \citenamefont {Reuter}}]{Daum:2009dn}%
  \BibitemOpen
  \bibfield  {author} {\bibinfo {author} {\bibfnamefont {J.-E.}\ \bibnamefont
  {Daum}}, \bibinfo {author} {\bibfnamefont {U.}~\bibnamefont {Harst}}, \ and\
  \bibinfo {author} {\bibfnamefont {M.}~\bibnamefont {Reuter}},\ }\href
  {\doibase 10.1007/JHEP01(2010)084} {\bibfield  {journal} {\bibinfo  {journal}
  {JHEP}\ }\textbf {\bibinfo {volume} {01}},\ \bibinfo {pages} {084} (\bibinfo
  {year} {2010})},\ \Eprint {http://arxiv.org/abs/0910.4938} {arXiv:0910.4938
  [hep-th]} \BibitemShut {NoStop}%
\bibitem [{\citenamefont {Daum}\ \emph {et~al.}(2011)\citenamefont {Daum},
  \citenamefont {Harst},\ and\ \citenamefont {Reuter}}]{Daum:2010bc}%
  \BibitemOpen
  \bibfield  {author} {\bibinfo {author} {\bibfnamefont {J.~E.}\ \bibnamefont
  {Daum}}, \bibinfo {author} {\bibfnamefont {U.}~\bibnamefont {Harst}}, \ and\
  \bibinfo {author} {\bibfnamefont {M.}~\bibnamefont {Reuter}},\ }\href
  {\doibase 10.1007/s10714-010-1032-2} {\bibfield  {journal} {\bibinfo
  {journal} {Gen. Rel. Grav.}\ }\textbf {\bibinfo {volume} {43}},\ \bibinfo
  {pages} {2393} (\bibinfo {year} {2011})},\ \Eprint
  {http://arxiv.org/abs/1005.1488} {arXiv:1005.1488 [hep-th]} \BibitemShut
  {NoStop}%
\bibitem [{\citenamefont {De~Brito}\ \emph
  {et~al.}(2019{\natexlab{b}})\citenamefont {De~Brito}, \citenamefont
  {Eichhorn},\ and\ \citenamefont {Pereira}}]{DeBrito:2019gdd}%
  \BibitemOpen
  \bibfield  {author} {\bibinfo {author} {\bibfnamefont {G.~P.}\ \bibnamefont
  {De~Brito}}, \bibinfo {author} {\bibfnamefont {A.}~\bibnamefont {Eichhorn}},
  \ and\ \bibinfo {author} {\bibfnamefont {A.~D.}\ \bibnamefont {Pereira}},\
  }\href {\doibase 10.1007/JHEP09(2019)100} {\bibfield  {journal} {\bibinfo
  {journal} {JHEP}\ }\textbf {\bibinfo {volume} {09}},\ \bibinfo {pages} {100}
  (\bibinfo {year} {2019}{\natexlab{b}})},\ \Eprint
  {http://arxiv.org/abs/1907.11173} {arXiv:1907.11173 [hep-th]} \BibitemShut
  {NoStop}%
\bibitem [{\citenamefont {Oda}\ and\ \citenamefont
  {Yamada}(2016)}]{Oda:2015sma}%
  \BibitemOpen
  \bibfield  {author} {\bibinfo {author} {\bibfnamefont {K.-y.}\ \bibnamefont
  {Oda}}\ and\ \bibinfo {author} {\bibfnamefont {M.}~\bibnamefont {Yamada}},\
  }\href {\doibase 10.1088/0264-9381/33/12/125011} {\bibfield  {journal}
  {\bibinfo  {journal} {Class. Quant. Grav.}\ }\textbf {\bibinfo {volume}
  {33}},\ \bibinfo {pages} {125011} (\bibinfo {year} {2016})},\ \Eprint
  {http://arxiv.org/abs/1510.03734} {arXiv:1510.03734 [hep-th]} \BibitemShut
  {NoStop}%
\bibitem [{\citenamefont {Eichhorn}\ \emph {et~al.}(2016)\citenamefont
  {Eichhorn}, \citenamefont {Held},\ and\ \citenamefont
  {Pawlowski}}]{Eichhorn:2016esv}%
  \BibitemOpen
  \bibfield  {author} {\bibinfo {author} {\bibfnamefont {A.}~\bibnamefont
  {Eichhorn}}, \bibinfo {author} {\bibfnamefont {A.}~\bibnamefont {Held}}, \
  and\ \bibinfo {author} {\bibfnamefont {J.~M.}\ \bibnamefont {Pawlowski}},\
  }\href {\doibase 10.1103/PhysRevD.94.104027} {\bibfield  {journal} {\bibinfo
  {journal} {Phys. Rev. D}\ }\textbf {\bibinfo {volume} {94}},\ \bibinfo
  {pages} {104027} (\bibinfo {year} {2016})},\ \Eprint
  {http://arxiv.org/abs/1604.02041} {arXiv:1604.02041 [hep-th]} \BibitemShut
  {NoStop}%
\bibitem [{\citenamefont {Hamada}\ and\ \citenamefont
  {Yamada}(2017)}]{Hamada:2017rvn}%
  \BibitemOpen
  \bibfield  {author} {\bibinfo {author} {\bibfnamefont {Y.}~\bibnamefont
  {Hamada}}\ and\ \bibinfo {author} {\bibfnamefont {M.}~\bibnamefont
  {Yamada}},\ }\href {\doibase 10.1007/JHEP08(2017)070} {\bibfield  {journal}
  {\bibinfo  {journal} {JHEP}\ }\textbf {\bibinfo {volume} {08}},\ \bibinfo
  {pages} {070} (\bibinfo {year} {2017})},\ \Eprint
  {http://arxiv.org/abs/1703.09033} {arXiv:1703.09033 [hep-th]} \BibitemShut
  {NoStop}%
\bibitem [{\citenamefont {Wetterich}(1993)}]{Wetterich:1992yh}%
  \BibitemOpen
  \bibfield  {author} {\bibinfo {author} {\bibfnamefont {C.}~\bibnamefont
  {Wetterich}},\ }\href {\doibase 10.1016/0370-2693(93)90726-X} {\bibfield
  {journal} {\bibinfo  {journal} {Phys. Lett. B}\ }\textbf {\bibinfo {volume}
  {301}},\ \bibinfo {pages} {90} (\bibinfo {year} {1993})},\ \Eprint
  {http://arxiv.org/abs/1710.05815} {arXiv:1710.05815 [hep-th]} \BibitemShut
  {NoStop}%
\bibitem [{\citenamefont {Morris}(1994)}]{Morris:1993qb}%
  \BibitemOpen
  \bibfield  {author} {\bibinfo {author} {\bibfnamefont {T.~R.}\ \bibnamefont
  {Morris}},\ }\href {\doibase 10.1142/S0217751X94000972} {\bibfield  {journal}
  {\bibinfo  {journal} {Int. J. Mod. Phys. A}\ }\textbf {\bibinfo {volume}
  {9}},\ \bibinfo {pages} {2411} (\bibinfo {year} {1994})},\ \Eprint
  {http://arxiv.org/abs/hep-ph/9308265} {arXiv:hep-ph/9308265} \BibitemShut
  {NoStop}%
\bibitem [{\citenamefont {Dupuis}\ \emph {et~al.}(2021)\citenamefont {Dupuis},
  \citenamefont {Canet}, \citenamefont {Eichhorn}, \citenamefont {Metzner},
  \citenamefont {Pawlowski}, \citenamefont {Tissier},\ and\ \citenamefont
  {Wschebor}}]{Dupuis:2020fhh}%
  \BibitemOpen
  \bibfield  {author} {\bibinfo {author} {\bibfnamefont {N.}~\bibnamefont
  {Dupuis}}, \bibinfo {author} {\bibfnamefont {L.}~\bibnamefont {Canet}},
  \bibinfo {author} {\bibfnamefont {A.}~\bibnamefont {Eichhorn}}, \bibinfo
  {author} {\bibfnamefont {W.}~\bibnamefont {Metzner}}, \bibinfo {author}
  {\bibfnamefont {J.~M.}\ \bibnamefont {Pawlowski}}, \bibinfo {author}
  {\bibfnamefont {M.}~\bibnamefont {Tissier}}, \ and\ \bibinfo {author}
  {\bibfnamefont {N.}~\bibnamefont {Wschebor}},\ }\href {\doibase
  10.1016/j.physrep.2021.01.001} {\bibfield  {journal} {\bibinfo  {journal}
  {Phys. Rept.}\ }\textbf {\bibinfo {volume} {910}},\ \bibinfo {pages} {1}
  (\bibinfo {year} {2021})},\ \Eprint {http://arxiv.org/abs/2006.04853}
  {arXiv:2006.04853 [cond-mat.stat-mech]} \BibitemShut {NoStop}%
\bibitem [{\citenamefont {Robinson}\ and\ \citenamefont
  {Wilczek}(2006)}]{Robinson:2005fj}%
  \BibitemOpen
  \bibfield  {author} {\bibinfo {author} {\bibfnamefont {S.~P.}\ \bibnamefont
  {Robinson}}\ and\ \bibinfo {author} {\bibfnamefont {F.}~\bibnamefont
  {Wilczek}},\ }\href {\doibase 10.1103/PhysRevLett.96.231601} {\bibfield
  {journal} {\bibinfo  {journal} {Phys. Rev. Lett.}\ }\textbf {\bibinfo
  {volume} {96}},\ \bibinfo {pages} {231601} (\bibinfo {year} {2006})},\
  \Eprint {http://arxiv.org/abs/hep-th/0509050} {arXiv:hep-th/0509050 [hep-th]}
  \BibitemShut {NoStop}%
\bibitem [{\citenamefont {Toms}(2007)}]{Toms:2007sk}%
  \BibitemOpen
  \bibfield  {author} {\bibinfo {author} {\bibfnamefont {D.~J.}\ \bibnamefont
  {Toms}},\ }\href {\doibase 10.1103/PhysRevD.76.045015} {\bibfield  {journal}
  {\bibinfo  {journal} {Phys. Rev. D}\ }\textbf {\bibinfo {volume} {76}},\
  \bibinfo {pages} {045015} (\bibinfo {year} {2007})},\ \Eprint
  {http://arxiv.org/abs/0708.2990} {arXiv:0708.2990 [hep-th]} \BibitemShut
  {NoStop}%
\bibitem [{\citenamefont {Ebert}\ \emph {et~al.}(2008)\citenamefont {Ebert},
  \citenamefont {Plefka},\ and\ \citenamefont {Rodigast}}]{Ebert:2007gf}%
  \BibitemOpen
  \bibfield  {author} {\bibinfo {author} {\bibfnamefont {D.}~\bibnamefont
  {Ebert}}, \bibinfo {author} {\bibfnamefont {J.}~\bibnamefont {Plefka}}, \
  and\ \bibinfo {author} {\bibfnamefont {A.}~\bibnamefont {Rodigast}},\ }\href
  {\doibase 10.1016/j.physletb.2008.01.037} {\bibfield  {journal} {\bibinfo
  {journal} {Phys. Lett. B}\ }\textbf {\bibinfo {volume} {660}},\ \bibinfo
  {pages} {579} (\bibinfo {year} {2008})},\ \Eprint
  {http://arxiv.org/abs/0710.1002} {arXiv:0710.1002 [hep-th]} \BibitemShut
  {NoStop}%
\bibitem [{\citenamefont {Anber}\ \emph {et~al.}(2011)\citenamefont {Anber},
  \citenamefont {Donoghue},\ and\ \citenamefont {El-Houssieny}}]{Anber:2010uj}%
  \BibitemOpen
  \bibfield  {author} {\bibinfo {author} {\bibfnamefont {M.~M.}\ \bibnamefont
  {Anber}}, \bibinfo {author} {\bibfnamefont {J.~F.}\ \bibnamefont {Donoghue}},
  \ and\ \bibinfo {author} {\bibfnamefont {M.}~\bibnamefont {El-Houssieny}},\
  }\href {\doibase 10.1103/PhysRevD.83.124003} {\bibfield  {journal} {\bibinfo
  {journal} {Phys. Rev. D}\ }\textbf {\bibinfo {volume} {83}},\ \bibinfo
  {pages} {124003} (\bibinfo {year} {2011})},\ \Eprint
  {http://arxiv.org/abs/1011.3229} {arXiv:1011.3229 [hep-th]} \BibitemShut
  {NoStop}%
\bibitem [{\citenamefont {Ellis}\ and\ \citenamefont
  {Mavromatos}(2012)}]{Ellis:2010rw}%
  \BibitemOpen
  \bibfield  {author} {\bibinfo {author} {\bibfnamefont {J.}~\bibnamefont
  {Ellis}}\ and\ \bibinfo {author} {\bibfnamefont {N.~E.}\ \bibnamefont
  {Mavromatos}},\ }\href {\doibase 10.1016/j.physletb.2012.04.005} {\bibfield
  {journal} {\bibinfo  {journal} {Phys. Lett. B}\ }\textbf {\bibinfo {volume}
  {711}},\ \bibinfo {pages} {139} (\bibinfo {year} {2012})},\ \Eprint
  {http://arxiv.org/abs/1012.4353} {arXiv:1012.4353 [hep-th]} \BibitemShut
  {NoStop}%
\bibitem [{\citenamefont {Toms}(2011)}]{Toms:2011zza}%
  \BibitemOpen
  \bibfield  {author} {\bibinfo {author} {\bibfnamefont {D.~J.}\ \bibnamefont
  {Toms}},\ }\href {\doibase 10.1103/PhysRevD.84.084016} {\bibfield  {journal}
  {\bibinfo  {journal} {Phys. Rev. D}\ }\textbf {\bibinfo {volume} {84}},\
  \bibinfo {pages} {084016} (\bibinfo {year} {2011})}\BibitemShut {NoStop}%
\bibitem [{\citenamefont {Felipe}\ \emph {et~al.}(2011)\citenamefont {Felipe},
  \citenamefont {Brito}, \citenamefont {Sampaio},\ and\ \citenamefont
  {Nemes}}]{Felipe:2011rs}%
  \BibitemOpen
  \bibfield  {author} {\bibinfo {author} {\bibfnamefont {J.~C.~C.}\
  \bibnamefont {Felipe}}, \bibinfo {author} {\bibfnamefont {L.~C.~T.}\
  \bibnamefont {Brito}}, \bibinfo {author} {\bibfnamefont {M.}~\bibnamefont
  {Sampaio}}, \ and\ \bibinfo {author} {\bibfnamefont {M.~C.}\ \bibnamefont
  {Nemes}},\ }\href {\doibase 10.1016/j.physletb.2011.04.061} {\bibfield
  {journal} {\bibinfo  {journal} {Phys. Lett. B}\ }\textbf {\bibinfo {volume}
  {700}},\ \bibinfo {pages} {86} (\bibinfo {year} {2011})},\ \Eprint
  {http://arxiv.org/abs/1103.5824} {arXiv:1103.5824 [hep-th]} \BibitemShut
  {NoStop}%
\bibitem [{\citenamefont {Narain}\ and\ \citenamefont
  {Anishetty}(2013)}]{Narain:2012te}%
  \BibitemOpen
  \bibfield  {author} {\bibinfo {author} {\bibfnamefont {G.}~\bibnamefont
  {Narain}}\ and\ \bibinfo {author} {\bibfnamefont {R.}~\bibnamefont
  {Anishetty}},\ }\href {\doibase 10.1007/JHEP07(2013)106} {\bibfield
  {journal} {\bibinfo  {journal} {JHEP}\ }\textbf {\bibinfo {volume} {07}},\
  \bibinfo {pages} {106} (\bibinfo {year} {2013})},\ \Eprint
  {http://arxiv.org/abs/1211.5040} {arXiv:1211.5040 [hep-th]} \BibitemShut
  {NoStop}%
\bibitem [{\citenamefont {de~Brito}\ and\ \citenamefont
  {Eichhorn}(2022{\natexlab{a}})}]{deBrito:2022vbr}%
  \BibitemOpen
  \bibfield  {author} {\bibinfo {author} {\bibfnamefont {G.~P.}\ \bibnamefont
  {de~Brito}}\ and\ \bibinfo {author} {\bibfnamefont {A.}~\bibnamefont
  {Eichhorn}},\ }\href@noop {} {\  (\bibinfo {year} {2022}{\natexlab{a}})},\
  \Eprint {http://arxiv.org/abs/2201.11402} {arXiv:2201.11402 [hep-th]}
  \BibitemShut {NoStop}%
\bibitem [{\citenamefont {Alkofer}\ \emph {et~al.}(2020)\citenamefont
  {Alkofer}, \citenamefont {Eichhorn}, \citenamefont {Held}, \citenamefont
  {Nieto}, \citenamefont {Percacci},\ and\ \citenamefont
  {Schr\"ofl}}]{Alkofer:2020vtb}%
  \BibitemOpen
  \bibfield  {author} {\bibinfo {author} {\bibfnamefont {R.}~\bibnamefont
  {Alkofer}}, \bibinfo {author} {\bibfnamefont {A.}~\bibnamefont {Eichhorn}},
  \bibinfo {author} {\bibfnamefont {A.}~\bibnamefont {Held}}, \bibinfo {author}
  {\bibfnamefont {C.~M.}\ \bibnamefont {Nieto}}, \bibinfo {author}
  {\bibfnamefont {R.}~\bibnamefont {Percacci}}, \ and\ \bibinfo {author}
  {\bibfnamefont {M.}~\bibnamefont {Schr\"ofl}},\ }\href {\doibase
  10.1016/j.aop.2020.168282} {\bibfield  {journal} {\bibinfo  {journal} {Annals
  Phys.}\ }\textbf {\bibinfo {volume} {421}},\ \bibinfo {pages} {168282}
  (\bibinfo {year} {2020})},\ \Eprint {http://arxiv.org/abs/2003.08401}
  {arXiv:2003.08401 [hep-ph]} \BibitemShut {NoStop}%
\bibitem [{\citenamefont {Held}(2019)}]{Held:2019vmi}%
  \BibitemOpen
  \bibfield  {author} {\bibinfo {author} {\bibfnamefont {A.}~\bibnamefont
  {Held}},\ }\emph {\bibinfo {title} {{From particle physics to black holes:
  The predictive power of asymptotic safety.}}},\ \href {\doibase
  10.11588/heidok.00027607} {Ph.D. thesis},\ \bibinfo  {school} {U. Heidelberg
  (main)} (\bibinfo {year} {2019})\BibitemShut {NoStop}%
\bibitem [{\citenamefont {Sirunyan}\ \emph
  {et~al.}(2018{\natexlab{a}})\citenamefont {Sirunyan} \emph
  {et~al.}}]{CMS:2018uxb}%
  \BibitemOpen
  \bibfield  {author} {\bibinfo {author} {\bibfnamefont {A.~M.}\ \bibnamefont
  {Sirunyan}} \emph {et~al.} (\bibinfo {collaboration} {CMS}),\ }\href
  {\doibase 10.1103/PhysRevLett.120.231801} {\bibfield  {journal} {\bibinfo
  {journal} {Phys. Rev. Lett.}\ }\textbf {\bibinfo {volume} {120}},\ \bibinfo
  {pages} {231801} (\bibinfo {year} {2018}{\natexlab{a}})},\ \Eprint
  {http://arxiv.org/abs/1804.02610} {arXiv:1804.02610 [hep-ex]} \BibitemShut
  {NoStop}%
\bibitem [{\citenamefont {Aaboud}\ \emph
  {et~al.}(2018{\natexlab{a}})\citenamefont {Aaboud} \emph
  {et~al.}}]{ATLAS:2018mme}%
  \BibitemOpen
  \bibfield  {author} {\bibinfo {author} {\bibfnamefont {M.}~\bibnamefont
  {Aaboud}} \emph {et~al.} (\bibinfo {collaboration} {ATLAS}),\ }\href
  {\doibase 10.1016/j.physletb.2018.07.035} {\bibfield  {journal} {\bibinfo
  {journal} {Phys. Lett. B}\ }\textbf {\bibinfo {volume} {784}},\ \bibinfo
  {pages} {173} (\bibinfo {year} {2018}{\natexlab{a}})},\ \Eprint
  {http://arxiv.org/abs/1806.00425} {arXiv:1806.00425 [hep-ex]} \BibitemShut
  {NoStop}%
\bibitem [{\citenamefont {Sirunyan}\ \emph
  {et~al.}(2018{\natexlab{b}})\citenamefont {Sirunyan} \emph
  {et~al.}}]{CMS:2018nsn}%
  \BibitemOpen
  \bibfield  {author} {\bibinfo {author} {\bibfnamefont {A.~M.}\ \bibnamefont
  {Sirunyan}} \emph {et~al.} (\bibinfo {collaboration} {CMS}),\ }\href
  {\doibase 10.1103/PhysRevLett.121.121801} {\bibfield  {journal} {\bibinfo
  {journal} {Phys. Rev. Lett.}\ }\textbf {\bibinfo {volume} {121}},\ \bibinfo
  {pages} {121801} (\bibinfo {year} {2018}{\natexlab{b}})},\ \Eprint
  {http://arxiv.org/abs/1808.08242} {arXiv:1808.08242 [hep-ex]} \BibitemShut
  {NoStop}%
\bibitem [{\citenamefont {Aaboud}\ \emph
  {et~al.}(2018{\natexlab{b}})\citenamefont {Aaboud} \emph
  {et~al.}}]{ATLAS:2018kot}%
  \BibitemOpen
  \bibfield  {author} {\bibinfo {author} {\bibfnamefont {M.}~\bibnamefont
  {Aaboud}} \emph {et~al.} (\bibinfo {collaboration} {ATLAS}),\ }\href
  {\doibase 10.1016/j.physletb.2018.09.013} {\bibfield  {journal} {\bibinfo
  {journal} {Phys. Lett. B}\ }\textbf {\bibinfo {volume} {786}},\ \bibinfo
  {pages} {59} (\bibinfo {year} {2018}{\natexlab{b}})},\ \Eprint
  {http://arxiv.org/abs/1808.08238} {arXiv:1808.08238 [hep-ex]} \BibitemShut
  {NoStop}%
\bibitem [{\citenamefont {Aad}\ \emph {et~al.}(2015)\citenamefont {Aad} \emph
  {et~al.}}]{ATLAS:2015xst}%
  \BibitemOpen
  \bibfield  {author} {\bibinfo {author} {\bibfnamefont {G.}~\bibnamefont
  {Aad}} \emph {et~al.} (\bibinfo {collaboration} {ATLAS}),\ }\href {\doibase
  10.1007/JHEP04(2015)117} {\bibfield  {journal} {\bibinfo  {journal} {JHEP}\
  }\textbf {\bibinfo {volume} {04}},\ \bibinfo {pages} {117} (\bibinfo {year}
  {2015})},\ \Eprint {http://arxiv.org/abs/1501.04943} {arXiv:1501.04943
  [hep-ex]} \BibitemShut {NoStop}%
\bibitem [{\citenamefont {Sirunyan}\ \emph
  {et~al.}(2018{\natexlab{c}})\citenamefont {Sirunyan} \emph
  {et~al.}}]{CMS:2017zyp}%
  \BibitemOpen
  \bibfield  {author} {\bibinfo {author} {\bibfnamefont {A.~M.}\ \bibnamefont
  {Sirunyan}} \emph {et~al.} (\bibinfo {collaboration} {CMS}),\ }\href
  {\doibase 10.1016/j.physletb.2018.02.004} {\bibfield  {journal} {\bibinfo
  {journal} {Phys. Lett. B}\ }\textbf {\bibinfo {volume} {779}},\ \bibinfo
  {pages} {283} (\bibinfo {year} {2018}{\natexlab{c}})},\ \Eprint
  {http://arxiv.org/abs/1708.00373} {arXiv:1708.00373 [hep-ex]} \BibitemShut
  {NoStop}%
\bibitem [{\citenamefont {Kowalska}\ \emph {et~al.}(2022)\citenamefont
  {Kowalska}, \citenamefont {Pramanick},\ and\ \citenamefont
  {Sessolo}}]{Kowalska:2022ypk}%
  \BibitemOpen
  \bibfield  {author} {\bibinfo {author} {\bibfnamefont {K.}~\bibnamefont
  {Kowalska}}, \bibinfo {author} {\bibfnamefont {S.}~\bibnamefont {Pramanick}},
  \ and\ \bibinfo {author} {\bibfnamefont {E.~M.}\ \bibnamefont {Sessolo}},\
  }\href@noop {} {\  (\bibinfo {year} {2022})},\ \Eprint
  {http://arxiv.org/abs/2204.00866} {arXiv:2204.00866 [hep-ph]} \BibitemShut
  {NoStop}%
\bibitem [{\citenamefont {Cai}\ \emph {et~al.}(2017)\citenamefont {Cai},
  \citenamefont {Herrero-Garc\'\i{}a}, \citenamefont {Schmidt}, \citenamefont
  {Vicente},\ and\ \citenamefont {Volkas}}]{Cai:2017jrq}%
  \BibitemOpen
  \bibfield  {author} {\bibinfo {author} {\bibfnamefont {Y.}~\bibnamefont
  {Cai}}, \bibinfo {author} {\bibfnamefont {J.}~\bibnamefont
  {Herrero-Garc\'\i{}a}}, \bibinfo {author} {\bibfnamefont {M.~A.}\
  \bibnamefont {Schmidt}}, \bibinfo {author} {\bibfnamefont {A.}~\bibnamefont
  {Vicente}}, \ and\ \bibinfo {author} {\bibfnamefont {R.~R.}\ \bibnamefont
  {Volkas}},\ }\href {\doibase 10.3389/fphy.2017.00063} {\bibfield  {journal}
  {\bibinfo  {journal} {Front. in Phys.}\ }\textbf {\bibinfo {volume} {5}},\
  \bibinfo {pages} {63} (\bibinfo {year} {2017})},\ \Eprint
  {http://arxiv.org/abs/1706.08524} {arXiv:1706.08524 [hep-ph]} \BibitemShut
  {NoStop}%
\bibitem [{\citenamefont {Klein}\ \emph {et~al.}(2019)\citenamefont {Klein},
  \citenamefont {Lindner},\ and\ \citenamefont {Ohmer}}]{Klein:2019iws}%
  \BibitemOpen
  \bibfield  {author} {\bibinfo {author} {\bibfnamefont {C.}~\bibnamefont
  {Klein}}, \bibinfo {author} {\bibfnamefont {M.}~\bibnamefont {Lindner}}, \
  and\ \bibinfo {author} {\bibfnamefont {S.}~\bibnamefont {Ohmer}},\ }\href
  {\doibase 10.1007/JHEP03(2019)018} {\bibfield  {journal} {\bibinfo  {journal}
  {JHEP}\ }\textbf {\bibinfo {volume} {03}},\ \bibinfo {pages} {018} (\bibinfo
  {year} {2019})},\ \Eprint {http://arxiv.org/abs/1901.03225} {arXiv:1901.03225
  [hep-ph]} \BibitemShut {NoStop}%
\bibitem [{\citenamefont {de~Brito}\ and\ \citenamefont
  {Eichhorn}(2022{\natexlab{b}})}]{deBrito:2022}%
  \BibitemOpen
  \bibfield  {author} {\bibinfo {author} {\bibfnamefont {G.~P.}\ \bibnamefont
  {de~Brito}}\ and\ \bibinfo {author} {\bibfnamefont {A.}~\bibnamefont
  {Eichhorn}},\ }\href@noop {} {\  (\bibinfo {year} {2022}{\natexlab{b}})},\
  \Eprint {http://arxiv.org/abs/2201.11402} {arXiv:2201.11402 [hep-th]}
  \BibitemShut {NoStop}%
\bibitem [{\citenamefont {Eichhorn}\ \emph
  {et~al.}(2021{\natexlab{b}})\citenamefont {Eichhorn}, \citenamefont {Pauly},\
  and\ \citenamefont {Ray}}]{Eichhorn:2021tsx}%
  \BibitemOpen
  \bibfield  {author} {\bibinfo {author} {\bibfnamefont {A.}~\bibnamefont
  {Eichhorn}}, \bibinfo {author} {\bibfnamefont {M.}~\bibnamefont {Pauly}}, \
  and\ \bibinfo {author} {\bibfnamefont {S.}~\bibnamefont {Ray}},\ }\href
  {\doibase 10.1007/JHEP10(2021)100} {\bibfield  {journal} {\bibinfo  {journal}
  {JHEP}\ }\textbf {\bibinfo {volume} {10}},\ \bibinfo {pages} {100} (\bibinfo
  {year} {2021}{\natexlab{b}})},\ \Eprint {http://arxiv.org/abs/2107.07949}
  {arXiv:2107.07949 [hep-ph]} \BibitemShut {NoStop}%
\end{thebibliography}%

\end{document}